\documentclass[lettersize,journal]{IEEEtran}
\usepackage{amsmath,amsfonts}
\usepackage{algorithmic}
\usepackage{algorithm}
\usepackage{array}
\usepackage[caption=false,font=normalsize,labelfont=sf,textfont=sf]{subfig}
\usepackage{textcomp}
\usepackage{stfloats}
\usepackage{url}
\usepackage{verbatim}
\usepackage{graphicx}
\usepackage{booktabs}
\usepackage{cite}
\usepackage{hyperref}
\usepackage{cleveref}
\usepackage{etoolbox}

\makeatletter
\def\@IEEEBIOskipN{0.8\baselineskip}

\patchcmd{\IEEEbiography}
  {\vskip \@IEEEBIOskipN plus 1fil minus 0\baselineskip}
  {\vskip \@IEEEBIOskipN}
  {}{}
\makeatother
\usepackage[table,svgnames,dvipsnames]{xcolor}
\usepackage{multirow}
\definecolor{revisioncolor}{HTML}{1C00FF}
\newcommand{\revise}[1]{#1}
\scriptsize
\renewcommand{\arraystretch}{1.12}
\usepackage{makecell}

\usepackage[svgnames]{xcolor}
\usepackage[normalem]{ulem}
\definecolor{myblue}{RGB}{114, 142, 253}
\definecolor{mypurple}{RGB}{128, 112, 181}
\definecolor{mypink}{RGB}{191, 128, 146}

\definecolor{vpink}{RGB}{193, 145, 165}
\newcommand{\pinksquare}{\kern-0.1em\tikz\fill[vpink] (0,0) rectangle (0.8em,0.8em);}

\definecolor{vyellow}{RGB}{214, 197, 165}
\newcommand{\yellowsquare}{\kern-0.1em\tikz\fill[vyellow] (0,0) rectangle (0.8em,0.8em);}

\definecolor{vblue}{RGB}{130, 150, 197}
\newcommand{\bluesquare}{\kern-0.1em\tikz\fill[vblue] (0,0) rectangle (0.8em,0.8em);}

\definecolor{vgray}{RGB}{108, 108, 107}
\newcommand{\grayring}{\tikz[baseline=-0.3em]\draw[vgray, line width=0.5pt] (0,0) circle (0.4em);}
\newcommand{\halfring}{\tikz[baseline=-0.1em]\draw[vgray, line width=2pt] (0.4em,0) arc (0:180:0.4em);}

\definecolor{vgreen}{RGB}{189, 191, 178}
\newcommand{\greensquare}{\kern-0.1em\tikz\fill[vgreen] (0,0) rectangle (0.8em,0.8em);}

\definecolor{vdarkgray}{RGB}{105, 105, 105}
\newcommand{\darkgraysquare}{%
  \kern-0.1em\tikz
  \fill[vdarkgray] (0,0) rectangle (0.8em,0.8em);
}

\definecolor{vpurple}{RGB}{184, 180, 203}
\newcommand{\purplesquare}{\kern-0.1em\tikz\fill[vpurple] (0,0) rectangle (0.8em,0.8em);}

\definecolor{vbluegray}{RGB}{248, 249, 250}
\newcommand{\bluegraysquare}{\kern-0.1em\tikz{\fill[vbluegray] (0,0) rectangle (0.8em,0.8em); \draw[gray, line width=0.3pt] (0,0) rectangle (0.8em,0.8em);}}

\definecolor{vpinkregion}{RGB}{229, 209, 216}
\newcommand{\pinkregion}{\kern-0.1em\tikz\fill[vpinkregion] (0,0) rectangle (0.8em,0.8em);}

\definecolor{vblueregion}{RGB}{203, 212, 229}
\newcommand{\blueregion}{\kern-0.1em\tikz\fill[vblueregion] (0,0) rectangle (0.8em,0.8em);}

\definecolor{vyellowregion}{RGB}{238, 230, 217}
\newcommand{\yellowregion}{\kern-0.1em\tikz\fill[vyellowregion] (0,0) rectangle (0.8em,0.8em);}

\newcommand{\bluedashedline}{\kern0.15em\tikz[baseline=0pt]{%
  \draw[vblue, line width=0.8pt, dash pattern=on 1.5pt off 1pt, rounded corners=0.5pt] (0,0) -- (0,0.82em) -- (0.51em,0.82em);%
  \draw[vblue, line width=0.8pt, dash pattern=on 1.5pt off 1pt, rounded corners=0.5pt] (1.42em,0) -- (1.42em,0.82em) -- (0.91em,0.82em);%
  \fill[vblue] (0.51em,0.63em) -- (0.91em,0.82em) -- (0.51em,1.07em) -- cycle;%
}}

\definecolor{vflatgray}{RGB}{206, 206, 206}
\newcommand{\flatgraybar}{\raisebox{0em}{\tikz[baseline=0pt]\fill[vflatgray] (0,0) rectangle (1.6em,0.35em);}}

\newcommand{\whitesquare}{\kern-0.1em\tikz{\fill[white] (0,0) rectangle (0.8em,0.8em); \draw[gray, line width=0.3pt] (0,0) rectangle (0.8em,0.8em);}}

\newcommand{\stripedbox}{\kern-0.1em\raisebox{0em}{\includegraphics[height=0.8em]{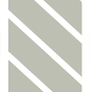}}}

\newcommand{\purplestripedbox}{\kern-0.1em\raisebox{0em}{\includegraphics[height=0.8em]{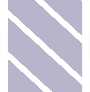}}}

\newcommand{\uparrowlegend}{\raisebox{-0.05em}{\includegraphics[height=0.55em]{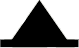}}}
\newcommand{\downarrowlegend}{\raisebox{-0.05em}{\includegraphics[height=0.55em]{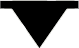}}}

\newcommand{\blueflow}{\raisebox{-0.15em}{\includegraphics[height=0.9em]{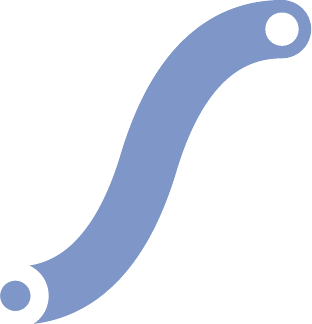}}}

\newcommand{\greendensity}{\raisebox{-0.15em}{\includegraphics[height=0.9em]{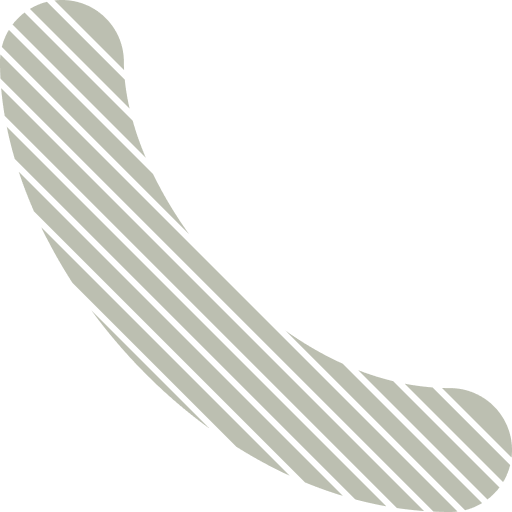}}}

\newcommand{\solidgradline}{\kern0.15em\raisebox{-0.05em}{\includegraphics[height=0.85em]{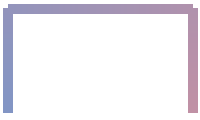}}}

\definecolor{vdarkgreen}{RGB}{130, 133, 110}
\newcommand{\greenpolyline}{\kern0.15em\tikz[baseline=0.1em]{%
  \draw[vdarkgreen, line width=1.2pt, line join=miter] (0,0.2em) -- (0.3em,0.45em) -- (0.6em,0.2em) -- (0.9em,0.45em);%
}}

\newcommand{\greenstaircase}{\kern0.15em\tikz[baseline=0.1em]{%
  \draw[vdarkgreen, line width=1.2pt, line join=miter] (0,0.2em) -- (0.4em,0.2em) -- (0.4em,0.55em) -- (0.8em,0.55em) -- (0.8em,0.2em) -- (1.2em,0.2em);%
}}

\newcommand{\purplepolyline}{\kern0.15em\tikz[baseline=0.1em]{%
  \draw[vpurple, line width=1.2pt, line join=miter] (0,0.2em) -- (0.3em,0.45em) -- (0.6em,0.2em) -- (0.9em,0.45em);%
}}

\newcommand{\purplestaircase}{\kern0.15em\tikz[baseline=0.1em]{%
  \draw[vpurple, line width=1.2pt, line join=miter] (0,0.2em) -- (0.4em,0.2em) -- (0.4em,0.55em) -- (0.8em,0.55em) -- (0.8em,0.2em) -- (1.2em,0.2em);%
}}

\newcommand{\framefive}{\raisebox{-0.15em}{\includegraphics[height=0.7em]{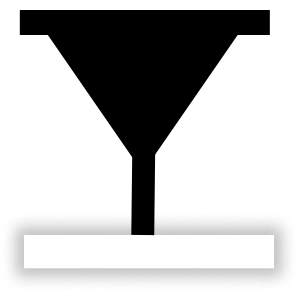}}}

\newcommand{\framesix}{\raisebox{-0.2em}{\includegraphics[height=0.85em]{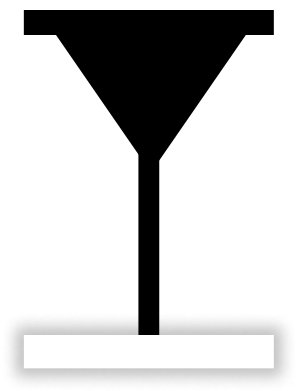}}}

\newcommand{\glyphlegend}{%
  \kern0.15em\raisebox{0em}{\tikz[baseline=-0.25em]{%
    \draw[vpink, line width=2.0pt, line cap=butt] (90:0.45em) arc (90:-30:0.45em);%
    \draw[vyellow, line width=2.0pt, line cap=butt] (210:0.45em) arc (210:90:0.45em);%
    \draw[vblue, line width=2.0pt, line cap=butt] (330:0.45em) arc (330:210:0.45em);%
  }}%
}

\usepackage{tikz}
\usepackage{amsmath,amssymb}
\newcommand{\techName}[1]{\textit{SAVVY}}

\newcommand{\SExpress}[1]{\textit{Query Expression}}
\newcommand{\SVerify}[1]{\textit{Query Verification}}
\newcommand{\SExecute}[1]{\textit{Query Execution}}
\newcommand{\SResult}[1]{\textit{Result Analysis}}

\definecolor{mydarkgray}{HTML}{626262}

\newcommand*\pic[1]{\tikz[baseline=(char.base)]{
\node[shape=rectangle, fill=mydarkgray, text=white, inner sep=1pt, minimum size=8pt, rounded corners=1.5pt] (char) {\textbf{#1}}}}

\definecolor{casepicborder}{HTML}{626262}
\definecolor{casepicfill}{HTML}{FFFFFF}

\newcommand*\casepic[1]{\tikz[baseline=(char.base)]{
\node[
  shape=rectangle,
  draw=casepicborder,
  fill=casepicfill,
  text=casepicborder,
  line width=1pt,
  inner xsep=1.2pt,
  inner ysep=1.2pt,
  minimum height=10.5pt,
  font=\bfseries\fontsize{9pt}{9pt}\selectfont,
  rounded corners=1.1pt
] (char) {\rlap{#1}\kern0.2pt #1}}}

\definecolor{ConceptColor}{RGB}{114, 142, 253}
\newcommand*\Concept[1]{\tikz[baseline=(char.base)]{
\node[shape=rectangle,fill=white, draw=ConceptColor, text=ConceptColor, inner sep= 1pt,minimum size=8pt,rounded corners=1.5pt] (char) {\textbf{#1}}}}

\definecolor{GroupColor}{RGB}{128, 112, 181}
\newcommand*\Group[1]{\tikz[baseline=(char.base)]{
\node[shape=rectangle,fill=white, draw=GroupColor, text=GroupColor, inner sep= 1pt,minimum size=8pt,rounded corners=1.5pt] (char) {\textbf{#1}}}}

\definecolor{IndividualColor}{RGB}{191, 128, 146}
\newcommand*\Individual[1]{\tikz[baseline=(char.base)]{
\node[shape=rectangle,fill=white, draw=IndividualColor, text=IndividualColor, inner sep= 1pt,minimum size=8pt,rounded corners=1.5pt] (char) {\textbf{#1}}}}

\definecolor{DataColor}{RGB}{114, 142, 253}

\definecolor{CaptionColor}{RGB}{89, 91, 97}

\begin{document}

\title{\techName{}: \underline{S}tudent \underline{A}ttention \underline{V}isualization for \underline{V}ideo-based Learning Anal\underline{y}sis}

\author{
Shixian Zhou, Minghuan Shen, Xiaolin Wen, Zijun Qiu, Yongliang Jiang,\\
Xiangyang Wu, Fei Wu, Yong Wang, and Zhiguang Zhou
\thanks{Shixian Zhou is with Hangzhou Dianzi University, Hangzhou, China. E-mail: shxzhou@hdu.edu.cn. He was a visiting student at Nanyang Technological University, Singapore during part of this work.}%
\thanks{Minghuan Shen, Zijun Qiu, Yongliang Jiang, Xiangyang Wu, and Zhiguang Zhou are with Hangzhou Dianzi University, Hangzhou, China. E-mail: \{251330014, 251330016, jiangyl, wuxy, zhgzhou\}@hdu.edu.cn.}%
\thanks{Xiaolin Wen, and Yong Wang are with Nanyang Technological University, Singapore. E-mail: \{xiaolin.wen, yong-wang\}@ntu.edu.sg}%
\thanks{Fei Wu is with Zhejiang University, Hangzhou, China. E-mail: wufei@zju.edu.cn}%
\thanks{(Corresponding authors: Yong Wang and Zhiguang Zhou.)}%
}


\markboth{Journal of \LaTeX\ Class Files,~Vol.~14, No.~8, August~2021}%
{Shell \MakeLowercase{\textit{et al.}}: A Sample Article Using IEEEtran.cls for IEEE Journals}


\maketitle
\begin{abstract}
Video-Based Learning (VBL) has become a popular delivery medium of education in the past decade, ranging from online education to hybrid learning. 
Students' rising expectations for video quality have motivated teachers to enhance the design of instructional videos before releasing them. 
Analyzing the attention of pilot cohorts in advance has become a conventional optimization strategy to guide course improvement. 
However, existing attention quantification algorithms are highly susceptible to noise in real-world environments, degrading estimation accuracy. 
Moreover, even when attention data are available, teachers must still invest substantial effort in empirical revision attempts, limiting practical feasibility. 
\revise{To address these challenges, we first propose a novel attention modeling framework based on multimodal brain signals that enables stable tracking of student attention levels.} 
We then develop \techName{}, a novel interactive visual analytics system that integrates visual and auditory attention to support top-down exploration of student attention variations.
\techName{} comprises three coordinated visualization modules.
These modules incorporate multi-level information, including course content structure, audiovisual information density, and attentional resource allocation, and provide multi-temporal-resolution attention trajectories of individual students, enabling teachers to comprehensively analyze the underlying causes of attention fluctuations and inform their subsequent instructional video improvement. 
We evaluate \techName{} through quantitative experiments, two case studies, and expert interviews. The results demonstrate the effectiveness and usability of \techName{} in intuitively identifying student attention variations and supporting instructional video optimization.
\end{abstract}
\section{Introduction} \label{sec:intro}
\IEEEPARstart{W}ith the rapid growth of online education, \textbf{\textit{Video-Based Learning (VBL)}} has gained significant momentum in recent years~\cite{sablic2021video}.
As the primary delivery medium for MOOCs~\cite{seaton2014does}, flipped classrooms~\cite{o2015use}, and hybrid learning~\cite{wieling2010impact}, instructional videos support students in constructing knowledge through visual and auditory channels~\cite{zhou2024conceptthread}.
The convenience of VBL has attracted a growing number of students and teachers, raising expectations for instructional video quality.
To improve teaching effectiveness, teachers commonly revise instructional videos prior to release~\cite{shi2015vismooc}, a process that supports pedagogical reflection and enhances the learning experience~\cite{sablic2021video}.
Consequently, reviewing and optimizing video content before release has become a standard step in VBL development workflows~\cite{weston1995model}.

\begin{figure*}[!htbp]
  \centering
  \setlength{\belowcaptionskip}{-0.6cm}
  \includegraphics[width=0.95\textwidth]{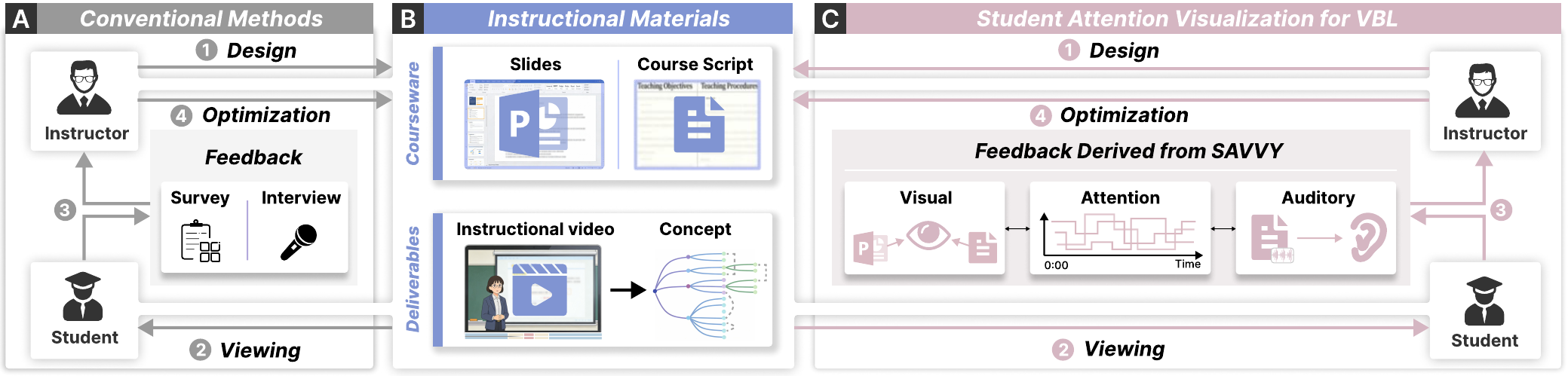}
  \caption{Comparison between conventional video revision approaches and ours. (A) Conventional approaches collect subjective feedback through questionnaires or interviews. (B) Teachers release instructional videos and revise them based on the collected feedback. (C) Our approach leverages student attention as objective feedback, tracing the causes of attention fluctuations from both visual and auditory dimensions to identify deficiencies in instructional design.}
  \label{fig:2}
\end{figure*}

However, video revision remains time-consuming and laborious due to the lack of effective support.
In practice, teachers often can only address superficial issues such as verbal slips~\cite{norman2017twelve}, which may fall short of students' expectations.
A conventional alternative is to collect student feedback in advance through closed pilot cohorts (e.g., surveys)~\cite{von2019iterative} (Fig.~\ref{fig:2}\pic{A}), but such feedback is inevitably affected by subjective bias.
Prior research suggests that attention, as a more objective indicator of student engagement, should serve as a core criterion when teachers revise instructional videos~\cite{fredricks2004school,brame2017effective}.
\revise{In this work, attention refers to students' engagement with the instructional video. We focus on the video-side factors that an instructor can control and that are known to shape students' attention and cognitive load, such as the amount of visual and auditory information~\cite{mayer2003nine}, slide composition~\cite{brame2017effective}, and narration pacing~\cite{ahn2025drives}. We do not model learner-side factors beyond an instructor's control, such as fatigue and mind-wandering~\cite{mcvay2012drifting}, the previous night's sleep~\cite{lim2010meta}, or prior familiarity with the topic~\cite{kalyuga2007expertise}.}
Using attention as feedback can make instructional video improvement more targeted and interpretable~\cite{chavan2022tcherly,deng2023review}.
However, the lack of real-time face-to-face interaction in VBL makes it difficult for teachers to directly assess students' attention~\cite{keegan1995distance}.

Therefore, recent studies have explored using electroencephalography (EEG) data from pilot cohorts to reflect students' attention states~\cite{rehman2025measuring,shaw2023attention,chen2017assessing}, thereby supporting course design improvement.
However, utilizing such EEG data to improve instructional videos remains challenging, because teachers still need to interpret attention dynamics with video content and reason about how different audiovisual presentation strategies influence learning engagement, which is particularly difficult for most teachers.

Visual analytics, with its strength in integrating multi-dimensional data and supporting interactive exploration, offers a promising approach for attention-based instructional video optimization.
By reviewing prior work on attention measurement and visualization~\cite{sabuncuoglu2023developing,aggarwal2021preliminary,duval2011attention,navarro2024vaad} and conducting a preliminary study (Sec.~\ref{sec:preliminary}) with four education experts, we identified two major challenges (\textbf{C1--C2}) in visualizing student attention for instructional video optimization.

\textbf{(C1) How to accurately quantify students' attention to support instructional video optimization in real-world settings?}
\revise{Attention is typically inferred from behavioral signals (e.g., posture~\cite{sabuncuoglu2023developing} and eye movements~\cite{sharma2016gaze}) using heuristic indicators (e.g., fixation duration and saccade ratio), which provide only coarse approximations and limited generalizability~\cite{deng2023review}, and struggle to capture attention fluctuations that occur within seconds~\cite{esterman2013zone}.}
As consumer-grade \textit{Brain-Computer Interfaces (BCI)} become widespread~\cite{xie2025eeg,aggarwal2021preliminary},
attention recognition models based on neural signals have achieved high accuracy in controlled settings~\cite{shaw2023attention}.
However, models relying on single-modality brain signals often face a trade-off between accuracy and stability in real-world settings~\cite{xu2022eeg}.
\revise{Therefore, there remains a critical need for methods that enable stable, accurate, and temporally fine-grained attention quantification for VBL in practice.}
\textbf{(C2) How to intuitively present the dynamic association patterns between course content and students' attention?}
In VBL, identifying optimization opportunities requires understanding how students' attention evolves in relation to course content, which is inherently complex.
Attention patterns vary across different instructional conditions, and multiple factors (e.g., inter-concept relationships and slide composition) can lead to distinct attention responses.
These factors often interact; for example, redundancy between slide text and teacher narration may affect attention distribution.
Furthermore, meaningful patterns require aggregating attention data from multiple students, making it increasingly challenging to analyze multi-factor attention variations as the student sample grows.
Existing attention visualization methods primarily support simple distribution representations~\cite{sauter2023behind,andrienko2012visual,navarro2024vaad}, which are inadequate for the multi-dimensional, multi-sample association analysis described above.
This creates a critical gap in enabling intuitive exploration of dynamic content-attention relationships in VBL.

To address these challenges, we propose \textbf{\techName{}}, an interactive visual analytics system that enables teachers to explore how audiovisual design in instructional videos affects student attention, thereby providing evidence for targeted optimization of instructional videos (Fig.~\ref{fig:2}\pic{C}).
Specifically, we first introduce a novel attention modeling framework that transforms complex multimodal neurophysiological signals into a sequence of temporally structured, attention-related feature vectors.
These features are then used to train a quantification model, enabling stable tracking of attention levels rather than relying on experimental conditions and qualitative indicators \textbf{(C1)}.
Further, we propose a novel visualization design comprising three coordinated modules linked by a shared timeline, which intuitively presents the dynamic association patterns between course content and students' attention through a top-down analytic workflow \textbf{(C2)}.
We quantitatively evaluate our proposed model and demonstrate its effectiveness in achieving accurate and stable attention quantification.
\revise{A sensitivity analysis further confirms that our content-derived metrics are robust to reasonable operationalizations.} We also conduct two case studies and expert interviews, showing that \techName{} effectively supports teachers in analyzing course-attention relationships and deriving actionable insights for instructional video optimization.
To the best of our knowledge, this is the first visual analytic system that leverages BCI to support VBL optimization.
Our main contributions are as follows:
\begin{itemize}
    \item Design requirements for visualizing students' attention through collaboration with four domain experts.
    \item A novel BCI-based attention modeling framework that analyzes students' attention by extracting attention cues from multi-modal neural signals.
    \item \techName{}, a visual analytics system that helps users comprehensively analyze students' attention patterns in VBL in a top-down manner.
    \revise{\item Quantitative evaluation of our model's performance and the robustness of our content-derived metrics, and qualitative validation of our system's effectiveness through two case studies and expert interviews with six experts.}
\end{itemize}

\section{Related Work}
This work is related to prior research on \textit{attention monitoring in learning} and \textit{visualization of students' attention}.

\subsection{Attention Monitoring in Learning}
Early research on attention monitoring in the learning process relies on students' subjective self-assessments~\cite{fredricks2004school,henrie2015measuring}.
Recently, intelligent algorithms for monitoring student attention have been proposed; depending on their input signals, these methods can be broadly categorized into behavior-based and physiology-based approaches~\cite{giannakos2019multimodal}.

\textbf{\textit{Behavior-based methods}} analyze attention by extracting students' behavioral manifestations in classroom settings, such as facial expressions, body posture~\cite{rudovic2019personalized,feng2025deep}, eye-movement trajectories~\cite{sharma2016gaze,madsen2021synchronized}, and weighted combinations of multiple behavioral cues~\cite{goldberg2021attentive,zhao2025research}.
These methods inherently rely on individual learning behaviors, making it difficult for the resulting models to generalize robustly.
\textbf{\textit{Physiology-based methods}} aim to analyze students' neurophysiological signals via BCIs to enable more stable attention monitoring in real-world scenarios.
Recent studies increasingly adopt machine learning pipelines, employing classifiers such as SVM~\cite{aggarwal2021preliminary,xie2025eeg}, KNN~\cite{asish2023internal}, and CNN-based deep networks~\cite{feng2025deep,nair2024human,devi2024ga} to perform attention state classification from neural signals.
In addition, Ur Rehman et al.~\cite{rehman2025measuring} propose a reinforcement learning approach to quantify attention from high-dimensional neural signals.
Nevertheless, these approaches predominantly rely on a single neurophysiological modality, making it difficult to circumvent the intrinsic limitations of individual brain-sensing signals (e.g., EEG's susceptibility to motion artifacts and fNIRS's hemodynamic latency), which often prevents models from reproducing laboratory-level performance in real-world settings.

In this work, our method extracts complementary features from multimodal brain signals under real-world conditions, thereby enabling robust quantification of students' attention.

\subsection{Visualization of students' attention}
Due to the inherent complexity of attention monitoring, a variety of visualization techniques have been proposed to support intuitive understanding of attentional changes~\cite{blascheck2014state,srinivasan2024attention}, including abstract depictions of eye-movement trajectories and gaze-focused highlighting~\cite{nguyen2015interactive}, as well as attention heatmaps~\cite{duchowski2015visualizing,chang2025tell}.
These techniques have been extensively adopted in domains such as commercial analytics and driving~\cite{jianu2025gaze}, yet only a limited body of work has explicitly focused on visualizing students' attention.
Most education-oriented studies rely primarily on attention heatmaps to assist teachers in optimizing course layout~\cite{conley2020examining} or to provide real-time awareness of students' attentional states for timely pedagogical interventions~\cite{rahman2020exploring,sauter2023behind}.
Building on this line of work, Hirzle et al.~\cite{hirzle2022attention} propose an embedded in-video region heatmap designed to proactively direct students' attention.
To enhance user friendliness, some systems favor more lightweight visual encodings such as bar charts and stacked plots to represent students' attention levels, thereby supporting downstream tasks such as course recommendation~\cite{duval2011attention} and self-assessment of engagement~\cite{sabuncuoglu2023developing}.
Recent work further combines attention heatmaps with gaze-point trajectory polylines and related graphical primitives~\cite{davalos2024gazeviz} to steer students' attention~\cite{thanyadit2023tutor} and to better differentiate attentional strategies across heterogeneous student cohorts~\cite{navarro2024vaad}.
However, these methods remain largely focused on localized comparisons of course content, designed to highlight students' current regions of interest rather than their actual levels of sustained attention.

In contrast, \techName{} takes students' actual attention as input and visualizes course content structure, audiovisual information density, and corresponding attentional resource allocation, helping teachers uncover students' attention variation patterns and thereby providing actionable evidence for instructional video optimization.

\section{Background}
This section introduces background information, including \textit{Neurophysiological Signals} and \textit{Double Deep Q-Network}.

\subsection{Neurophysiological Signals} \label{sec:Sig}
\textbf{Neurophysiological Signals} are physiological measurements reflecting nervous system activity, widely applied across various domains, including education and clinical medicine.
Common modalities include EEG, fNIRS, and EMG, each capturing different aspects of neural activity.
Since our work focuses on student attention monitoring, we adopt \textit{\textbf{EEG}} and \textit{\textbf{fNIRS}}, two mature modalities widely used for attention quantification~\cite{liu2024eeg}.
During preprocessing, fNIRS recordings are converted to hemoglobin concentration changes (HbO and HbR) via the modified Beer--Lambert law~\cite{fazli2012enhanced}, and both signals are band-pass filtered to retain cognitive-relevant frequency components, with noisy channels corrected through interpolation or excluded via ICA.

\revise{\textbf{EEG \& fNIRS}. They have complementary strengths and weaknesses. EEG offers millisecond-level temporal resolution that is sensitive to rapid stimulus changes~\cite{michel2012towards}, but it is susceptible to motion artifacts. And its spatial resolution is intrinsically limited because the poorly conducting skull spatially low-pass filters the cortical field~\cite{srinivasan1998spatial,nunez2006electric} and EEG source reconstruction is an ill-posed inverse problem~\cite{grech2008review}. In contrast, although fNIRS suffers from temporal latency due to the slow hemodynamic response, it directly measures cortical hemodynamics beneath each optode, provides region-level spatial specificity~\cite{ferrari2012brief}, and can offer greater tolerance to moderate movement in wearable settings. 

In our bimodal design, EEG primarily contributes high-temporal-resolution neural dynamics, whereas fNIRS provides complementary region-level spatial information over the targeted cortical regions. These complementary properties motivate our EEG--fNIRS bimodal input.}
Throughout this paper, neurophysiological signals refer specifically to bimodal EEG and fNIRS signals.

\subsection{Double Deep Q-Network} \label{sec: motif}
We quantify attention using a \textit{\textbf{Double Deep Q-Network (DDQN)}}, as deep reinforcement learning is well suited for modeling complex temporal dependencies in high-dimensional neurophysiological signals~\cite{yu2021reinforcement}, and its reinforcement mechanism has been shown to effectively extract attention levels from continuous brain signal inputs~\cite{mnih2015human}.
We formulate real-world student attention monitoring as a \textit{Markov decision process (MDP)}, as illustrated in Fig.~\ref{fig:framework}\pic{B1}.
\revise{Multimodal neurophysiological signals serve as the state input to the \textit{environment}, where informative representations including wavelet coefficients and spectral power estimates are derived to construct the \textit{state space}.}
Upon receiving a new batch of attention-related features, the \textit{environment} transitions to a new state.
The \textit{action space} of the \textit{agent} comprises three attention levels: high, medium, and low.
At each time step, the \textit{agent} selects an action based on its estimated \textit{Q-values}, and the selected action serves as the predicted attention state for that step.
Correct predictions receive a positive \textit{reward} signal, while incorrect ones incur a penalty.
Through this \textit{exploration-exploitation} and \textit{reward} mechanism, the \textit{agent} iteratively refines its attention quantification by updating its \textit{Q-network}.

\section{Informing the design}
This section presents our preliminary study and the design requirements derived from it.

\subsection{Preliminary Study}\label{sec:preliminary}
We conducted a preliminary study to comprehensively gather users' requirements regarding attention visualization for VBL optimization.
The details of the participants and procedures are as follows:

\textbf{Participants:}
We invited four domain experts in online education (\textbf{E1--E4}) as participants (three male and one female, age $= 41.8 \pm 6.5$).
\textbf{E1} and \textbf{E2} are professors from two different research universities, each with over 8 years of online teaching experience.
\textbf{E3} is an educational psychologist from a national institute of education.
\textbf{E4} is a postdoctoral researcher in Learning Sciences with a strong background in neuroscience and over 4 years of experience in student attention research, focusing particularly on optimizing instructional strategies based on attention feedback.

\textbf{Procedures:}
To ensure that our methodology aligns with participants' daily workflows, the preliminary study was conducted in two sessions.
In the first session, we conducted one-hour, one-on-one, semi-structured interviews with each participant.
Participants were first asked to describe the primary directions for revising instructional videos, including pacing and slide presentation, and to outline their expectations for visualization-based explanations.
We then posed a set of questions centered on how to optimize instructional videos based on attention cues.
Participants were encouraged to describe any challenges they had encountered or anticipated when applying these methods to address the needs they had previously articulated.
In response to these challenges, we distilled an initial set of design requirements and proposed a student attention modeling framework based on multimodal neural signals.
In the second session, participants were invited to evaluate the design requirements and the proposed framework, with a preliminary visual design mockup provided as a reference.
Their feedback guided us in finalizing the design requirements for instructional video optimization and refining the attention modeling framework.

The challenges (\textbf{C1}--\textbf{C2}) introduced in Section~\ref{sec:intro} were distilled from those raised by the participants during this study.
The design requirements and the attention quantification framework are presented in Section~\ref{sec:design_requirement} and Section~\ref{sec:Quantification}, respectively, and the interactive visual analytics system is described in Section~\ref{sec:framework}.

\subsection{Design Requirements}\label{sec:design_requirement}
\revise{During the preliminary study, all experts (\textbf{E1--E4}) pointed out that when optimizing an instructional video, course instructors need to determine when and where the course should be improved. Therefore, the core task is to locate when and why student attention declines and to determine the content that needs revision. This task decomposes into three core objects of analysis, each addressing a different part of the task: the \textit{\textbf{course concept}}, to locate the instructional content where attention loss occurs; the \textit{\textbf{student group}}, to analyze how group attention changes over time and attribute the fluctuations to the audiovisual content; and the \textit{\textbf{individual student}}, to verify the specific causes of attention loss in detail. This decomposition is also consistent with the top-down workflow agreed upon by the experts (\textbf{E1--E4}), which proceeds from the overall course structure to the specific causes of attention fluctuations.
Here and throughout the paper, a \textit{concept} refers to a relatively independent, self-contained knowledge unit in the course content, consistent with the definition in ConceptThread~\cite{zhou2024conceptthread}. For example, the neurons-and-synapses video discussed in Sec~\ref{sec:Brain Dataset} contains concepts such as the action potential, synaptic transmission, and neurotransmitters, and the convolutional-neural-network video contains concepts such as the convolution kernel, the stride, and the feature map.

Therefore, we categorize the six design requirements we derived (\textbf{R1--R6}) by these three analysis objects: the \Concept{Concept} category analyzes attention from the course structure, supporting rapid screening and preliminary attribution; the \Group{Group} category focuses on temporal attention trends and audiovisual instructional design; and the \Individual{Individual} category centers on each student's detailed attention trajectory and audiovisual attention allocation.}

\begin{itemize}
  \item{\textbf{R1}} \Concept{Concept} \textbf{Overview the course content structure.} All experts (\textbf{E1--E4}) agreed that concepts form the structural backbone of a course, interconnected through various relationships such as subsumption, derivation, and analogy, which together constitute the logical flow of the course. Accurately extracting concepts and their relationships is therefore a prerequisite for helping teachers review the overall course structure.

  \item{\textbf{R2}} \Concept{Concept} \textbf{Profile attention and audiovisual features by concept.} All experts (\textbf{E1--E4}) agreed that aggregating attention distribution by concept helps quickly locate segments with low engagement. \textbf{E2} stressed that concept importance, reflected by occurrence frequency and duration, should be presented alongside to help teachers prioritize revision targets. \textbf{E3} and \textbf{E4} noted that presenting the average audiovisual information density and corresponding attention resource allocation for each concept helps teachers preliminarily identify potential causes of low attention.

  \item{\textbf{R3}} \Group{Group} \textbf{Show temporal trends of group attention.} All experts (\textbf{E1--E4}) emphasized that attention decline often concentrates in specific periods within a concept, making it crucial to identify these critical moments. Therefore, our visualization should present the dynamic changes of group attention along the timeline, including the proportional distribution of different attention levels across periods.

  \item{\textbf{R4}} \Group{Group} \textbf{Support audiovisual diagnosis of instructional design.} Three experts (\textbf{E1, E3, E4}) mentioned that audiovisual information density in instructional videos varies with each slide transition, and that presenting group audiovisual information density and corresponding attention resource allocation per slide is key to diagnosing instructional design deficiencies. \textbf{E1} further noted that the image area ratio in slides and the similarity between the narration and slide text can additionally assist teachers in assessing the appropriateness of audiovisual design.

  \item{\textbf{R5}} \Individual{Individual} \textbf{Provide attention trajectories for individual students.} Three experts (\textbf{E1--E3}) noted that group aggregated results may mask individual differences, as attention patterns observed at the group level may not hold consistently across individual students. It is therefore necessary to let teachers inspect the attention trajectory of each student. \textbf{E2} further suggested that the trajectories should support switching between multiple temporal resolutions, enabling teachers to flexibly navigate between an individual's overall trends and local details.

  \item{\textbf{R6}} \Individual{Individual} \textbf{Display individual students' audiovisual attention allocation.} Three experts (\textbf{E1--E3}) mentioned that group audiovisual attention resource allocation can also mask individual differences. For example, a student with an extremely high visual channel proportion and one with an extremely low proportion can average out to a moderate value. Therefore, our visualization should present each student's audiovisual attention resource allocation to help teachers gain a deeper understanding of the channel composition of attention at the individual level.

\end{itemize}

\section{Data Processing}\label{sec:Data Processing}

\begin{figure}[!t]
  \centering
  \setlength{\belowcaptionskip}{-0.6cm}
  \includegraphics[width=\columnwidth]{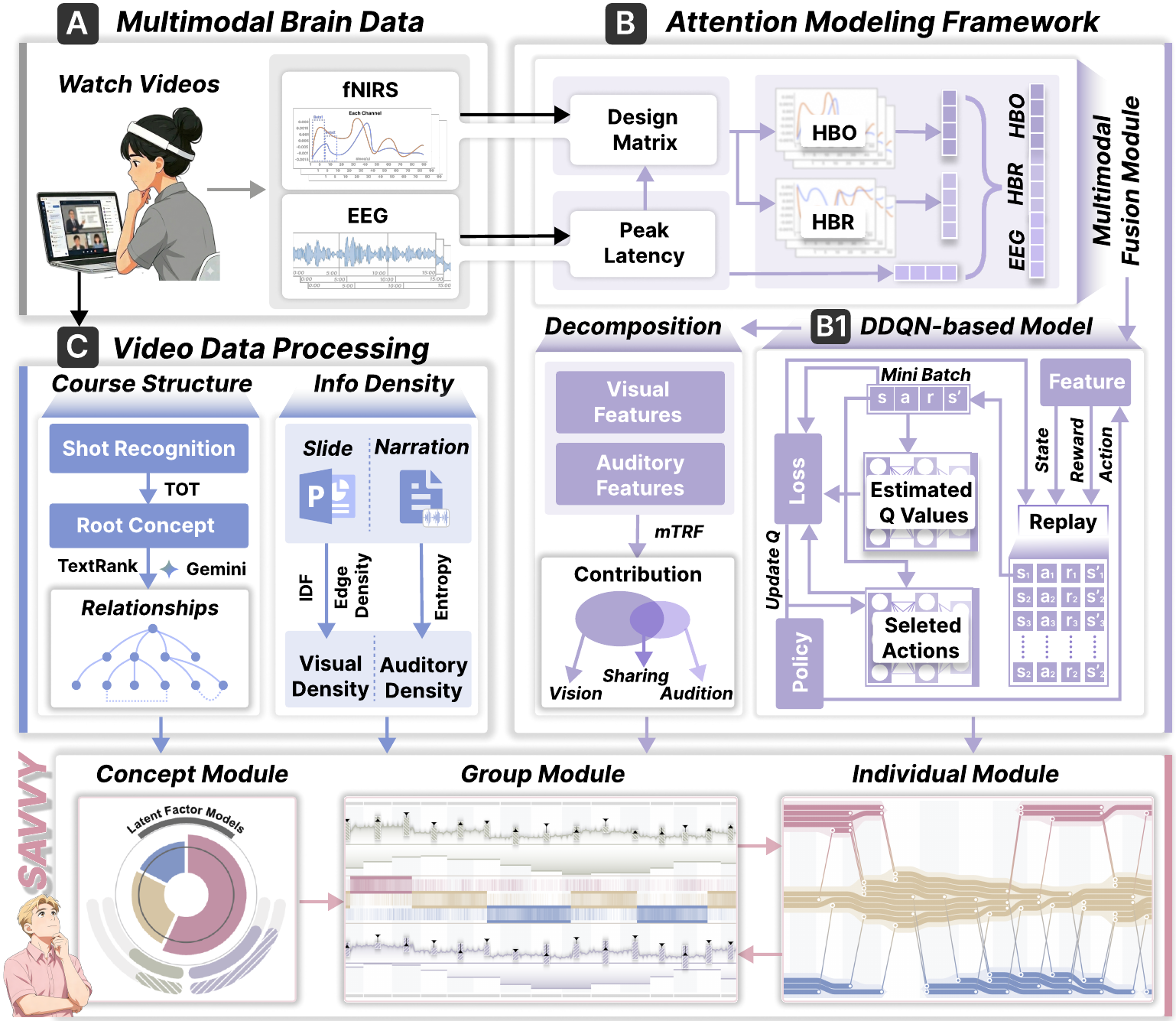}
  \caption{The data processing pipeline of \techName{}: (A) brain data acquisition and preprocessing, (B) attention modeling framework with a fusion module, a DDQN-based quantification model (B1), and an attention decomposition module, (C) course structure and information density extraction. These outputs feed into the three visualization modules of \techName{}.}
  \label{fig:framework}
\end{figure}

In this section, we present the data processing workflow of \techName{} (Fig.~\ref{fig:framework}).
We first collected an EEG--fNIRS dataset from students across three course sessions and preprocessed the signals following standard neuroscience protocols (Fig.~\ref{fig:framework}\pic{A}).
We then constructed an attention modeling framework comprising a multimodal fusion module, a attention quantification model, and an audiovisual attention decomposition module (Fig.~\ref{fig:framework}\pic{B}).
Finally, we analyzed the instructional videos, including course content structure extraction and audiovisual information density computation (Fig.~\ref{fig:framework}\pic{C}).
Together, these steps constitute the complete workflow of \techName{}, enabling teachers to gain deeper insights into attention variation patterns.

\subsection{Brain Dataset}\label{sec:Brain Dataset}

We collected a multimodal brain dataset with simultaneous EEG and fNIRS signals from 31 healthy participants (16 male, 15 female; mean age $23.6 \pm 1.2$ years) during instructional videos on \textit{convolutional neural networks}, \textit{matrix factorization}, and \textit{neurons and synapses}. 
\revise{All participants had normal vision and hearing and were well-rested, having slept adequately the night before. The instructional content was entirely new to them.}
\revise{Our sample of 31 participants exceeds the sample sizes typically used in BCI experiments, which commonly recruit around 20 participants and often fewer in BCI work~\cite{larson2017sample}, providing a broader empirical basis for constructing reliable attention labels.}
The experiment was IRB-approved (No.~2026052), and informed consent was obtained from all participants.

\textbf{Brain signals} were captured using the \textit{\textbf{LoongX}} Neurophysiological Headband~\cite{zhao2025wearable}, \revise{a lightweight wearable device that needs no conductive gel or setup preparation and is well suited to real-world deployment}. It integrates 2 EEG channels (125~Hz) and 8 fNIRS channels (25~Hz), \revise{placed over prefrontal and temporal regions that together subserve attention and audiovisual cognitive processing. In the temporal lobe, particularly the superior temporal sulcus, auditory, visual, and audiovisual signals converge and are integrated into unified percepts~\cite{beauchamp2004integration}. Making a perceptual decision over this integrated information then recruits top-down attentional control in the prefrontal cortex~\cite{corbetta2002control,gregoriou2009high,noppeney2021}. 
We therefore positioned the channels over these regions to capture the neural signals most relevant to attention and audiovisual cognitive processing.
Here, ``top-down'' refers to goal-directed attentional control in cognitive neuroscience. This usage is distinct from the term top-down workflow, which denotes the overview-to-detail visual analysis process in this paper.

We also deliberately avoided the occipital lobe, where the primary visual cortex performs low-level, stimulus-driven unimodal encoding~\cite{grill2004human}, preventing such early visual responses from contaminating our recordings. On the auditory side, the primary auditory cortex is buried deep within the lateral sulcus and is thus relatively inaccessible to surface EEG and fNIRS~\cite{warrier2009relating}.
Therefore, our temporal channels mainly reflect superficial temporal association activity rather than stimulus-driven auditory encoding, indirectly avoiding primary auditory processing as well. However, this layout can only bias the measured data toward neural activity of top-down control and multimodal integration, and cannot completely separate them.}

The dataset contains over \textbf{\textit{4.5 million}} data points, with each experiment lasting $50.2 \pm 7.6$ minutes. For each session, we preprocessed the signals following the procedure described in Section~\ref{sec:Sig} and applied a 0.1--50~Hz band-pass filter to preserve attention-relevant frequency bands.

\subsection{Experimental procedure}
\revise{The experiment had two phases, with EEG and fNIRS recorded throughout. \textbf{\textit{In the first phase}}, participants watched the instructional videos while external stimuli induced three attentional states (high, medium, and low), following a well-established paradigm in which controlled stimuli reliably elicit distinct cognitive and attentional states~\cite{shatil2014novel,esterman2013zone} and attention labels are built from such stimulus-induced states in BCI research~\cite{berka2007eeg,aci2019distinguishing,rehman2025measuring}. 
Three levels are chosen to provide richer information than a binary scheme while avoiding the blurred class boundaries and accuracy loss that more levels would cause~\cite{wan2021frontal}.
The three states were induced in a fixed order by graded manipulations of the viewing condition, such as playback speed, interpolated quizzes, screen brightness, and volume, to which participants responded by watching and answering the quizzes. Each recording was segmented with a sliding window, and every segment took the attention level its stimulus condition was designed to induce, yielding the initial design-based labels. \textbf{\textit{In the second phase}}, on the next day, participants freely watched a new video without stimulation or quizzes, and these data were reserved for case analysis.}

\revise{We then verified these labels after data collection. Two domain experts (\textbf{E3--E4}) independently judged whether each segment's initial attention-level label was correct, using three complementary indicators: quiz accuracy and response time~\cite{esterman2013zone,mcvay2012drifting}, the EEG spectrogram~\cite{berka2007eeg,aci2019distinguishing}, and the fNIRS-measured cortical activation~\cite{fishburn2014sensitivity}. 
When the experts agreed, the common label was used. 
When they disagreed, a co-author discussed the segment with them to reach the final label, and any segment that remained ambiguous after this discussion was discarded. The controlled induction together with this verification establishes reliable ground-truth labels. The full protocol, indicators, criteria, and supporting literature are detailed in the Supplementary Material.} The resulting labels (high, medium, and low attention) were used to train the attention quantification model. The dataset and its detailed description are available at \url{https://vis-savvy.github.io/SAVVY/}.

\subsection{Attention Modeling}\label{sec:Quantification}

\textbf{Multimodal Fusion.}
For EEG, we extracted spectral features via Welch's method and time--frequency features using the discrete wavelet transform.
For fNIRS, we used EEG-derived priors (peak amplitude and latency) to fit a GLM~\cite{gao2023hybrid} on HbO and HbR signals and then extracted CSP features.
Finally, we constructed a fused feature set $Z = \{z^{(1)}, z^{(2)}, \ldots, z^{(s)}\}$, where $z^{(i)} = [f_{\mathrm{EEG}}^{(i)};\; f_{\mathrm{fNIRS}}^{(i)}]$.

\textbf{Attention Quantification Model.}
We developed a DDQN-based attention quantification model. Using the labeled dataset collected in the first phase of the experiment, we split the data by participant into 80\% training and 20\% testing sets.
The fused features serve as the state $s_t$ and are fed into the DDQN \textit{agent}, which outputs an action $a_t$ corresponding to an attention level via the learned action value function $Q(s_t, a)$.
Action selection adopts a hybrid strategy combining Softmax and $\varepsilon$-greedy, maintaining sufficient exploration in early training while gradually decaying $\varepsilon$ to improve decision stability.
The ground-truth label $\hat{y}_t$ and the predicted label $y_t$ are passed to the environment to generate the \textit{reward} signal $r_t$.
To enhance training robustness, we integrate the \textit{\textbf{reward}} function with \textit{action confidence}:
{
\begin{equation}
r_t = \left(\alpha\,\mathbb{I}(\hat{y}_t = y_t) - \beta\,(1 - \mathbb{I}(\hat{y}_t = y_t))\right)c_t .
\end{equation}
}
Here, $\mathbb{I}(\hat{y}_t = y_t) \in \{0,1\}$ is an indicator function, $\alpha$ and $\beta$ are the reward and penalty coefficients for correct and incorrect predictions respectively, and $c_t \in [0,1]$ denotes the action confidence that weights the reward magnitude.
A parameter search yielded optimal accuracy at $\alpha=10$ and $\beta=1$.
This asymmetric design provides sufficient negative feedback to suppress repeated incorrect actions during training.

Transitions $(s_t, a_t, r_t, s_{t+1})$ are stored in a replay buffer, from which mini-batches are randomly sampled to reduce temporal correlations.
We further extend the standard DDQN loss with an episode-level term to better capture overall attention dynamics.
\revise{After training, the optimized model is used to quantify student attention on the test data, producing a predicted attention level (high, medium, or low) at a one-second temporal resolution. 
This predicted level is defined relative to our validated ground-truth labels in the machine-learning and BCI sense, rather than as a direct measurement of a participant’s internal attentional state in the psychological sense.}
Model training and deployment details, including the network architecture, hyperparameters, hardware, and training curves, are provided in the Supplementary Material.

\textbf{Audiovisual Attention Decomposition.} Following Desai et al.~\cite{desai2021generalizable}, we compute attentional resource allocation across visual and auditory channels via forward encoding and variance partitioning. \revise{For each attention segment output by our quantification model, we extract temporally aligned auditory and visual features. We then fit a multi-time-lag linear encoding model to estimate the variance in the neural data explained by each modality. Next, we apply variance partitioning to separate this explained variance into three components: the \textit{Visual Attention Contribution} (VAC), the \textit{Auditory Attention Contribution} (AAC), and the \textit{Shared Audiovisual Attention Contribution} (SAC). These three components are empirical surrogate indicators of the relative involvement of visual, auditory, and shared audiovisual processing. They do not directly measure attention as an internal state. Instead, they provide auxiliary cues for teachers’ exploratory analysis.}

\subsection{Instructional Video Processing}
\textbf{Course Structure.}
Effective course structure analysis requires identifying concepts and their relationships.
Following the pipeline in ConceptThread~\cite{zhou2024conceptthread}, we extracted root concepts and concept relationships, parsed slide hierarchies, and identified every slide transition.
We replaced the original LLM with Gemini-3-pro-preview to improve relationship extraction quality.

\begin{figure*}[!htbp]
  \centering
  \setlength{\belowcaptionskip}{-0.6cm}
  \includegraphics[width=0.95\textwidth]{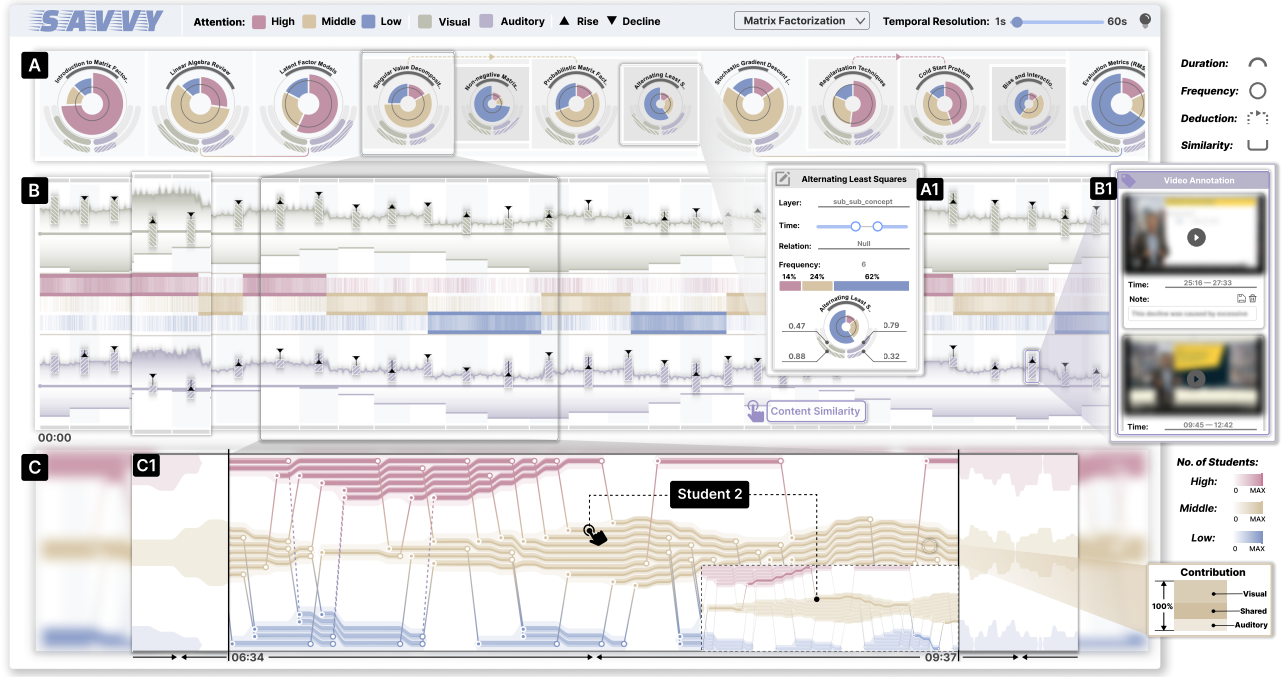}
  {
  \caption{The interface of \techName{} consists of the \textit{Concept Module} (A), the \textit{Group Module} (B), and the \textit{Individual Module} (C). The \textit{Concept Module} (A) presents the course content structure and detailed information of each concept (A1). The \textit{Group Module} (B) displays temporal trends of group attention and detailed audiovisual information, where the video annotation panel (B1) further assists teachers in locating and annotating video segments that require revision. The \textit{Individual Module} (C) provides attention change trajectories for each student (C1) to verify findings during the exploration process.}
  \label{fig:teaser}
  }
\end{figure*}

\textbf{Information Density.}\label{sec: info_density}
\revise{Grounded in the Limited Capacity Model of Motivated Mediated Message Processing (LC4MP)~\cite{fisher2018limited}, we treat slide-transition-induced event segments as our units of analysis and quantify the information density of both the visual and auditory channels.}
\revise{Here, visual and auditory information density are content-derived proxy indicators. They approximate the amount of visual and auditory information in the materials presented to learners. They describe the presented content itself, not learners' internal cognitive states, nor the amount of information each learner actually and selectively takes in.}

\revise{For the \textit{\textbf{visual channel}}, we partition each slide into image and text regions, using Mask Region-based Convolutional Neural Network (Mask R-CNN)~\cite{he2017mask} to detect the image regions and Differentiable Binarization Network (DBNet)~\cite{liao2020real} to detect the text regions.}
\revise{For the image region, visual complexity is an effective proxy for its information content~\cite{attneave1954some, donderi2006visual}, and edge density is widely used to measure visual complexity~\cite{rosenholtz2007measuring, miniukovich2014quantification}. Since instructional graphics are mostly structured diagrams composed of lines, boundaries, arrows, grids, and contours, edge density is particularly accurate at quantifying their visual complexity. We therefore adopt edge density to quantify the visual information content of the image region. Given the binary edge map $E_e$ and the image-region mask $M_F$,}
\revise{\begin{equation}
ED_e = \frac{\sum_{p \in \Omega} M_F(p)\, E_e(p)}{\sum_{p \in \Omega} M_F(p) + \epsilon},
\end{equation}}
\revise{where $\Omega$ is the set of slide pixels and $\epsilon$ avoids division by zero; we then normalize $ED_e$ to $\widehat{ED}_e$.}
\revise{For the text region, a term's IDF surprisal corresponds to its self-information~\cite{shannon1948mathematical} and admits a well-established information-theoretic interpretation, so the average IDF surprisal is an effective proxy for textual information content. Since slides present key points as phrases rather than continuous narration, we measure the slide text by the average normalized IDF of its tokens. For segment $e$ with $n_e$ tokens,}
\revise{\begin{equation}
S_e = \frac{1}{n_e} \sum_{k=1}^{n_e} s(w_{e,k}),
\end{equation}}
\revise{where $s(w)$ is the normalized smoothed IDF of term $w$ over the $N$ course slides.}
\revise{Finally, we combine the two regions into an overall measure. Since the spatial extent of on-screen content is widely used to quantify the visual complexity of displays and interfaces~\cite{reinecke2013}, and the relative area of image and text regions is a common feature when analyzing slides~\cite{shi2019, ferguson2017}, we weight each region's contribution by its area ratio. Let $\rho_F$ and $\rho_T$ be the image and text area ratios, i.e., each region's pixel count divided by the total slide area. These weights are set automatically by region coverage rather than by hand, and are not renormalized to sum to one, so that large blank areas lower the overall density. For segment $e$ with duration $\Delta t_e$, the visual information density is}
\revise{\begin{equation}
D^{\mathrm{vis}}_e = \frac{\rho_F \, \widehat{ED}_e + \rho_T \, S_e}{\Delta t_e}.
\end{equation}}

\revise{For the \textit{\textbf{auditory channel}}, Shannon entropy is the canonical measure of the information content of a source~\cite{shannon1948mathematical}, and the entropy of a word distribution in particular reflects the lexical diversity and information richness of language~\cite{shannon1951prediction, bentz2017entropy}. We therefore transcribe teacher speech into word sequences and use the entropy of its word distribution as a proxy for the information content of the narration. We choose entropy over IDF surprisal because teacher narration is continuous speech with frequent repetitions, for which IDF overlooks global distributional characteristics. We normalize this entropy and divide it by the narration duration to obtain the auditory information density. Additionally, we compute the semantic similarity between slide text and the aligned speech transcript within each unit to capture cross-channel content redundancy.}

\begin{figure*}[!htbp]
  \centering 
  \setlength{\belowcaptionskip}{-0.6cm}
  \includegraphics[width=0.95\textwidth
  ]{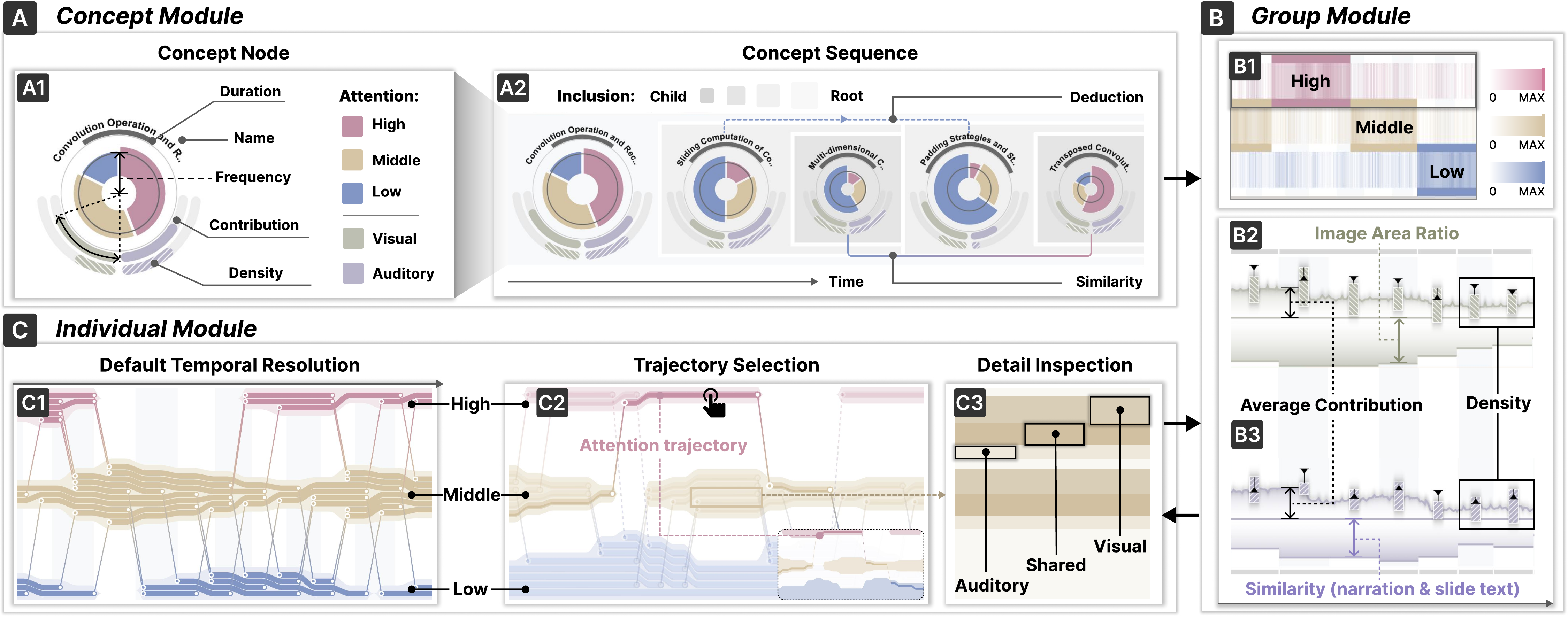}
  \caption{The visual design of \techName{}. The \textit{Concept Module} (A) presents the course content structure. The \textit{Group Module} (B) displays group-level attention variation trends (B1), visual-channel information (B2), and auditory-channel information (B3). The \textit{Individual Module} (C) provides detailed attention trajectories of individual students (C1) and fine-grained inspection of audiovisual attention contributions (C2).}
  \label{fig:visual_design}
\end{figure*}

\section{\techName{}}\label{sec:framework}
In this section, we present \techName{}, an interactive visual analytics system designed to help teachers intuitively explore patterns of student attention variation.
Fig.~\ref{fig:teaser} shows the interface of \techName{}, comprising three coordinated modules: the \textit{Concept Module} (A), the \textit{Group Module} (B), and the \textit{Individual Module} (C).
The \textit{Concept Module }(Fig.~\ref{fig:teaser}\pic{A}) presents the course content structure, concept-level attention distribution, and an overview of attention resource allocation.
The \textit{Group Module} (Fig.~\ref{fig:teaser}\pic{B}) visualizes group-level attention variation trends on a timeline and supports comparative analysis between audiovisual content and attention resource allocation.
The \textit{Individual Module} (Fig.~\ref{fig:teaser}\pic{C}) provides individual students' attention trajectories at multiple temporal resolutions (Fig.~\ref{fig:teaser}\pic{C1}).
With \techName{}, teachers can first click on concepts in the \textit{Concept Module} to identify target segments, then delve into the \textit{Group Module} to analyze associations between instructional design and attention fluctuations or brush time windows of interest, and verify the identified patterns with the support of the \textit{Individual Module}.

\subsection{Concept Module} 
\revise{As Nesbit et al.~\cite{nesbit2006learning} summarized, the concept map is a widely recognized and mature way to represent knowledge structure in learning, and nodes are a core component of concept maps. Therefore, \techName{} provides a \textit{Concept Module} (Fig.~\ref{fig:visual_design}\pic{A}) that helps teachers quickly screen concept segments for improvement and preliminarily attribute attention deviation, through two components: concept nodes (Fig.~\ref{fig:visual_design}\pic{A1}) and a concept sequence (Fig.~\ref{fig:visual_design}\pic{A2}) composed of these nodes.}

The \textit{\textbf{concept node}} displays the attention distribution, importance, and audiovisual features of each concept (\textbf{R2}).
Specifically, each concept node adopts a ``flower-shaped'' glyph design, as shown in Fig.~\ref{fig:visual_design}\pic{A1}.
Inside the glyph, we use a segmented ring~\glyphlegend{} to represent the distribution of three attention levels across all students when studying each concept.
The color of each segment encodes a different attention level: pink \pinksquare{} indicates high attention, yellow \yellowsquare{} indicates moderate attention, and blue \bluesquare{} indicates low attention.
\revise{The angle and height of each segment jointly encode the proportion of the current concept's lecture duration in the corresponding attention level. This design draws on redundant encoding~\cite{nothelfer2017redundant, chun2017redundant}. When two proportions are too close to distinguish by angle, it reveals the difference through the height.}

In addition, we overlay a dark-gray thin ring \grayring{} on top of the segmented ring, whose radius encodes the occurrence frequency of the concept.
Above the thin ring, we add an arc indicator \halfring{} in the same color, whose angle within the upper semicircle encodes the relative lecture duration of the concept.
Together, frequency and duration reflect the importance of each concept.
A curved concept name label is displayed above the arc indicator.
Through this design, teachers can quickly grasp the importance of each concept and the overall attention distribution of students.
In the lower half of the glyph, we use four curved progress bars to present audiovisual information.
The two bars on the left are colored dark green \greensquare{}, representing the visual channel, while the two on the right are colored purple~\purplesquare{}, representing the auditory channel.
The progress bars covered by white stripes represent information density (Section~\ref{sec: info_density}), while the uncovered progress bars represent the attention contribution (Section~\ref{sec:Quantification}) of the corresponding channel.
The angle occupied by each progress bar encodes the magnitude of the respective value.
Through this design, teachers can intuitively compare the audiovisual information density and attention contribution across concepts.

The \textit{\textbf{concept sequence}} displays the relationships among concepts (\textbf{R1}).
Specifically, the visual encoding of each relationship type is shown in Fig.~\ref{fig:visual_design}\pic{A2}.
For sequential relationships, we arrange concept nodes horizontally in teaching order, which forms the base layout of the concept sequence.
For subsumption relationships, we encode parent–child structures through hierarchical nesting. Parent nodes visually contain their children within light gray backgrounds. As the hierarchy deepens, node background colors become progressively darker and node sizes decrease, further reinforcing the hierarchy.
Derivation relationships are encoded as dashed lines~\bluedashedline{} above concept nodes, indicating that one concept is derived from another.
Analogy relationships are encoded as solid lines~\solidgradline{} below concept nodes, indicating that multiple concepts share similar conclusions or purposes.
The color of each line is determined by the dominant attention level of the concept at each end, with a gradient transition from one side to the other to reveal the potential influence of inter-concept relationships on attention.

\begin{figure}[!t]
  \centering
  \setlength{\belowcaptionskip}{-0.6cm}
  \includegraphics[width=0.85\columnwidth]{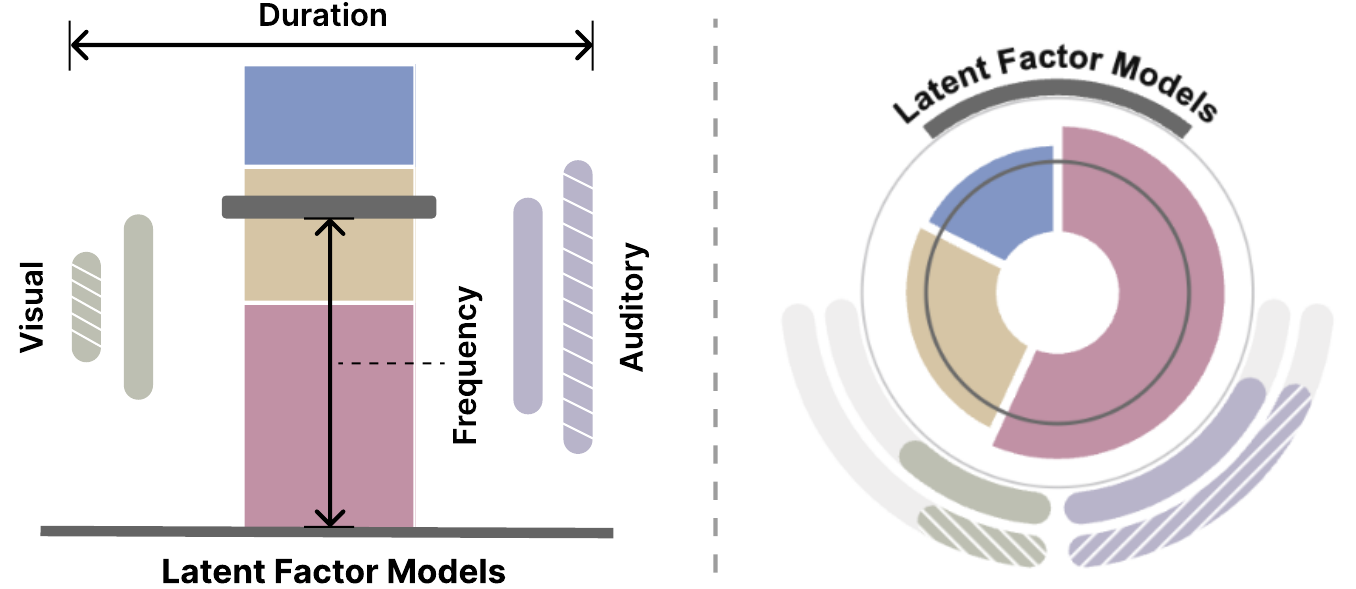}
  \caption{Comparison of an alternative design (left) and our design (right) for the concept node. The alternative uses bar charts to present the information teachers need for preliminary screening. Our design encodes the same information within a single glyph, while preserving the inter-concept relationships.}
  \label{fig:Alternative Design 1}
\end{figure}

\revise{\textbf{Design Alternatives.} Fig.~\ref{fig:Alternative Design 1} (left) presents an alternative bar-based glyph. It encodes the attention distribution with a central stacked bar, the occurrence frequency with the height of a dark-gray~\darkgraysquare{}bar above the x-axis, the audiovisual information with the lengths of the side bars, and the lecture duration with the gap between the two outermost information-density bars. However, this design proved inadequate. For a long lecture duration, the gap between the two information-density bars stretches and wastes visual space, and the loosely arranged bars do not form a cohesive unit. Moreover, when two attention levels occupy similar proportions, the stacked bar is hard to read. Its fixed height also prevents the glyph from shrinking at deeper levels, weakening the subsumption hierarchy. In contrast, our proposed design resolves these problems.}

\textbf{Interaction.} Clicking on a concept node synchronizes the \textit{Group Module} and \textit{Individual Module} to the corresponding time period. Double-clicking opens the editing panel (Fig.~\ref{fig:teaser}\pic{A1}) for adjusting lecture duration or modifying inter-concept relationships.

\subsection{Group Module}
The \textit{Group Module }(Fig.~\ref{fig:visual_design}\pic{B}) is a multi-track temporal design that reveals the dynamics of group attention and audiovisual course content over time (\textbf{R3, R4}). 
\revise{It is the diagnostic view after the \textit{Concept Module} overview, where teachers attribute an attention decline to specific audiovisual content. We therefore adopt the established design of stacking multiple tracks on a shared timeline for multivariate time-oriented data~\cite{brehmer2016timelines}, which aligns these heterogeneous streams in time.} 
It consists of three vertically stacked tracks sharing a common timeline: the attention track (Fig.~\ref{fig:visual_design}\pic{B1}), the visual track (Fig.~\ref{fig:visual_design}\pic{B2}), and the auditory track (Fig.~\ref{fig:visual_design}\pic{B3}). Horizontal bars~\flatgraybar{} along the upper and lower boundaries represent slides, while white-gray striped backgrounds indicate content changes within slides, enabling direct linkage between attention dynamics and slide transitions.

The \textit{\textbf{attention track}} (Fig.~\ref{fig:visual_design}\pic{B1}) adopts a barcode-based design to display group attention distribution over time (\textbf{R3}).
The y-axis shows three attention levels, and the x-axis represents time. 
\revise{Cell color encodes the number of students at each level per second, using a white-to-color gradient of the three attention colors defined in the \textit{Concept Module}, enabling teachers to quickly grasp the distribution across levels.}
To highlight attention trends, we mark the dominant attention level (i.e., the level with the most students) at each timestep, using symmetric bold outlines in the color of the dominant level above and below the barcodes.
To avoid frequent switching across levels, we introduce an attention smoothing annotation algorithm that preserves the current mark when short-term fluctuations occur, presenting continuous, stable trends.
Specifically, we determine the dominant state at each second by majority voting across all individual attention sequences, yielding a group-level dominant state sequence $x_t$:
{
\begin{equation}
x_t = \arg\max_{c \in \mathcal{S}} \sum_{i=1}^{N} \mathbb{I}\!\left(s_t^{(i)} = c\right),
\end{equation}
}
where $N$ denotes the total number of students and $\mathbb{I}(\cdot)$ is the indicator function.
We then apply a forward smoothing scan with a 10-second tolerance window: when a state change is detected, the algorithm does not immediately confirm a transition but searches forward for up to 10 seconds.
If the baseline state reappears, the intermediate deviations are treated as temporary fluctuations and merged into the current interval; otherwise, a genuine transition is confirmed.
Finally, any interval shorter than 5 seconds is treated as an unstable residual and merged into the preceding interval.
Through this track, teachers can intuitively identify periods of sustained attention decline.

The \textit{\textbf{visual track}} (Fig.~\ref{fig:visual_design}\pic{B2}) shows how visual stimuli affect group attention (\textbf{R4}), using a color scheme consistent with the dark green~\greensquare{} of the visual channel in the concept node.
Specifically, a polyline~\greenpolyline{} above the x-axis encodes temporal changes in VAC, while the heights of white-striped rectangles~\stripedbox{} along the polyline~\greenpolyline{} represent visual information density at each slide transition.
Triangles inside or outside each rectangle encode density changes between consecutive slides, with upward triangles~\uparrowlegend{} indicating increases and downward triangles~\downarrowlegend{} indicating decreases. 
The length of the connecting line encodes the magnitude of change.
The staircase line~\greenstaircase{} below the x-axis shares the same timeline, with its vertical position indicating the image area ratio at each slide transition.
Through this track, teachers can establish the connection between visual content and VAC.

The \textit{\textbf{auditory track}} (Fig.~\ref{fig:visual_design}\pic{B3}) shows how auditory stimuli influence group attention (\textbf{R4}), using a color scheme consistent with the purple~\purplesquare{} of the auditory channel. It follows the same design as the visual track, using a polyline~\purplepolyline{}, white-striped rectangles~\purplestripedbox{}, and triangles~\uparrowlegend{} to encode AAC, auditory information density, and density changes.
\revise{The key difference lies in the staircase line~\purplestaircase{} below the x-axis, where vertical position encodes the semantic similarity between the teacher's narration and slide text, the auditory counterpart to the visual track's image area ratio. This symmetric design better matches visual intuition and helps teachers read the two channels' different indicators in a unified way.}

\begin{figure}[!t]
  \centering
  \setlength{\belowcaptionskip}{-0.6cm}
  \includegraphics[width=0.95\columnwidth]{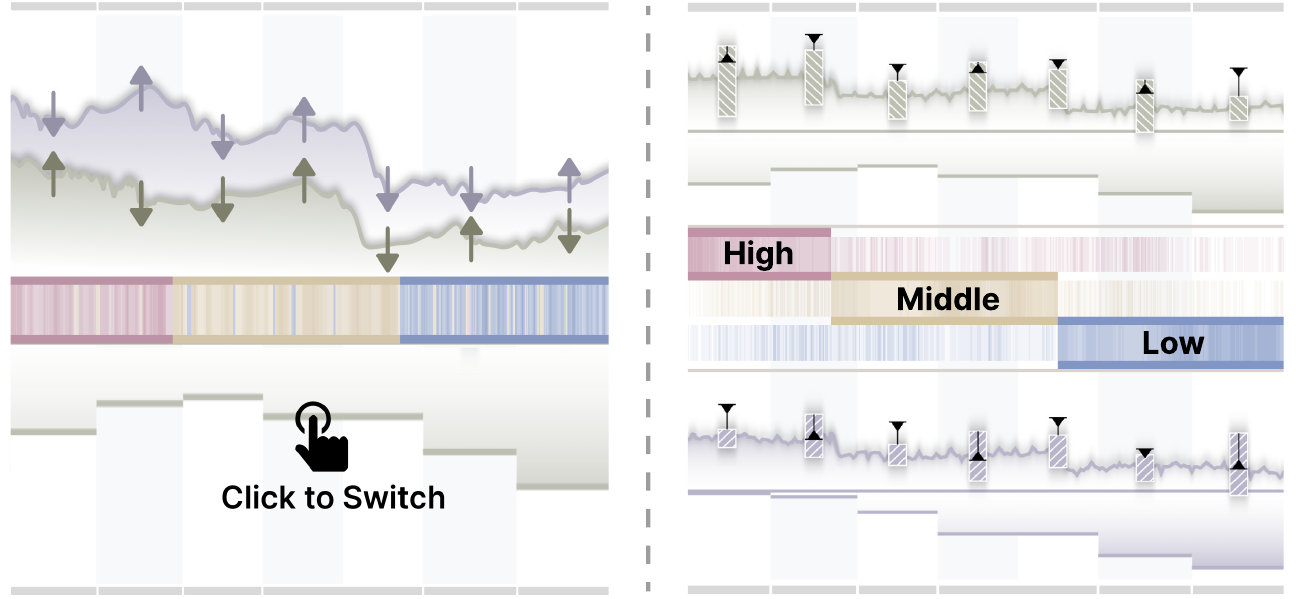}
  \caption{Comparison of an alternative design (left) and our design (right) for the \textit{Group Module}. The alternative uses more aggregated time-series visualizations to summarize attention and the audiovisual content over time. Our design separates each channel into an aligned track on a shared timeline, preserving all the data needed for the analysis.}
  \label{fig:Alternative Design 2}
\end{figure}

\revise{\textbf{Design Alternatives.} During the design process, we considered an alternative built from simpler time-series visualizations (Fig.~\ref{fig:Alternative Design 2}, left): a single heatmap row for the dominant attention level, a stacked area chart with arrows for information-density change above, and a click-switchable factor track below. However, the stacked area chart makes VAC and AAC hard to read, and its arrows give only the direction of change, so the analysis stays shallow. The heatmap row is cluttered and switches labels over very short spans, obscuring the trend. And clicking to switch factors prevents comparing a channel's attention with its factor at once. In contrast, our proposed design resolves these problems.}

\textbf{Interaction.} Teachers can brush the attention track to reset the focused time period across all modules. Density rectangles in the audiovisual tracks can be dynamically repositioned onto the x-axis to reveal occluded VAC and AAC polylines. \revise{Hovering over the visual or auditory track reveals a label naming the quantity it encodes, so teachers can tell the two tracks apart.} \revise{Clicking a density rectangle pins the original video clip for its exact time interval to the annotation panel on the right (Fig.~\ref{fig:teaser}\pic{B1}). By selecting several rectangles, teachers retrieve the actual content at those moments, play the clips in sync for side-by-side comparison, and verify it against the metrics \techName{} reports for the same interval. Teachers can annotate any clip to mark the content to be revised, and the panel can be collapsed when not needed.}

\subsection{Individual Module}
The \textit{Individual Module} is a streamgraph-based design that displays each student's attention details over time (\textbf{R5}).
As shown in Fig.~\ref{fig:visual_design}\pic{C}, we use a path along the timeline to visualize each student's \textit{attention change trajectory}.
The x-axis represents time, with resolution dynamically adjusted via user zoom interactions, where the striped background shows slide transitions consistent with the \textit{Group Module}.
Inspired by NFTDisk's flow-based design~\cite{wen2023nftdisk}, paths are grouped along the y-axis into three attention levels (high, moderate, low).
Paths within each group are enclosed side-by-side by a flow band whose vertical width encodes the student count at that level, like a streamgraph (Fig.~\ref{fig:visual_design}\pic{C1}).
The high-attention band is aligned to the upper boundary, while the low-attention band to the lower.
\revise{Path color encodes attention level following the same scheme as the \textit{Concept Module}, while transitions between levels are shown using linear gradient colors between the corresponding levels, with solid lines for single-level transitions and dashed lines for two-level transitions.}
This design reveals temporal changes in student count at each level through flow bands, highlights transitions between levels through connections between flow bands, and enables tracking individual changes through paths.

\textbf{Interaction.} The \textit{Individual Module} supports dynamic temporal resolution switching (\textbf{R5}), defaulting to coarse-grained trajectories aggregated at 60-second intervals. The number of partitions is determined by concept selections in the \textit{Concept Module}. Teachers can drag the locator line outward to zoom in, refining granularity down to 1 second, or inward to zoom out, gradually restoring to the default  resolution. Dragging the locator line in any partition causes all partitions to zoom proportionally, enabling fair comparison across time periods. When the resolution is finer than 30 seconds, two overlapping rectangles appear on each trajectory (Fig.~\ref{fig:visual_design}\pic{C3}), representing VAC and AAC, with the overlap representing SAC (\textbf{R6}). Independent portions are rendered lighter and the overlap darker. Teachers can also click on individual trajectories to show or hide them (Fig.~\ref{fig:visual_design}\pic{C2}), and the \textit{Group Module} will synchronously update to reflect the selected student's data.

\begin{figure*}[!htbp]
  \centering 
  \setlength{\belowcaptionskip}{-0.6cm}
  \includegraphics[width=\textwidth
  ]{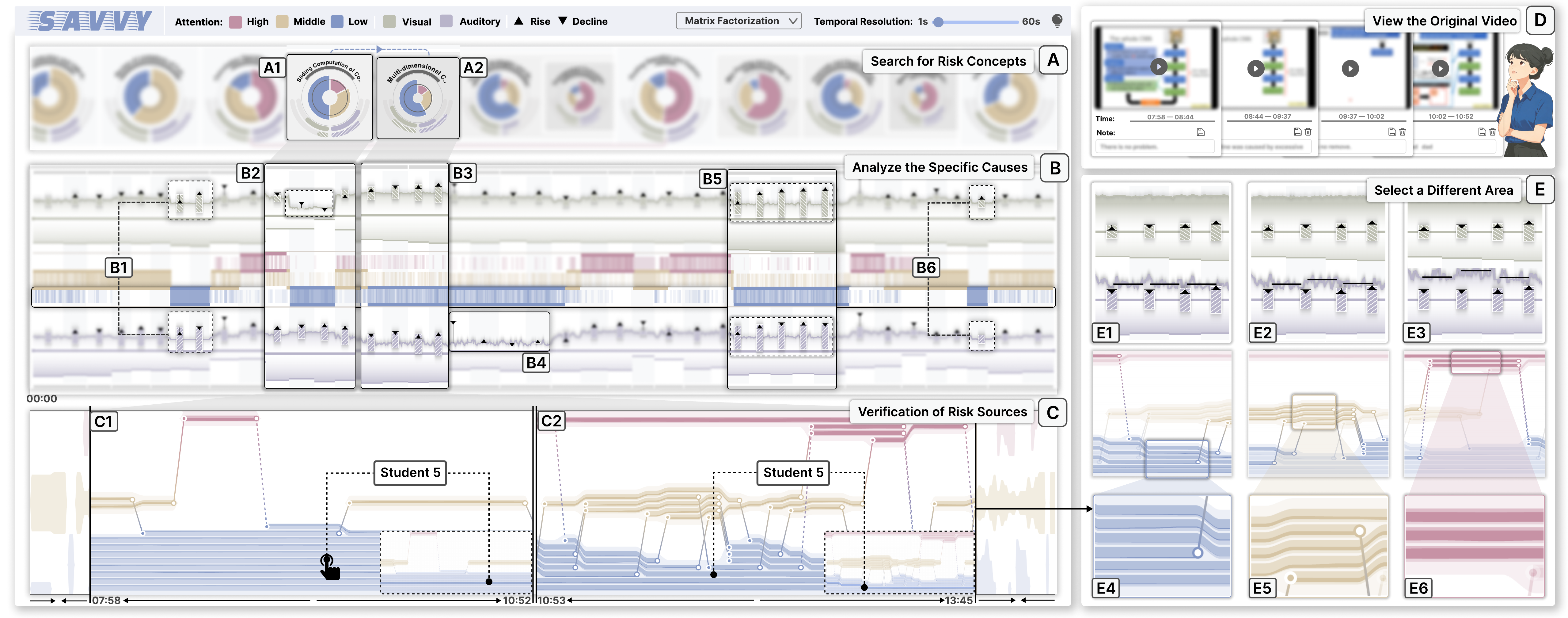}
  \caption{A teacher uses \techName{} to identify the causes of low attention. (A) The \textit{Concept Module} displays an overview of the course structure and helps the teacher quickly screen concepts for revision (A1-A2). (B) In the \textit{Group Module}, the teacher discovers anomalous information density and attention contributions (B1-B6). (C) In the \textit{Individual Module}, the teacher inspects attention details at different temporal resolutions (C1-C2) and selectively views individual students' attention details (E4-E6) by region (E1-E3), confirming the causes of attention decline.(D) The teacher traces attention changes back to the source material to verify the actual visual information density across presentation stages.}
  \label{fig:case1}
\end{figure*}
\section{Evaluation}
In our evaluation, we invited the experts from the preliminary study (\textbf{E1--E4}) along with two additional experts (\textbf{E5--E6}) to validate the usability and effectiveness of \techName{} through expert interviews. \textbf{E5} and \textbf{E6} are professors from different universities who specialize in applying brain-computer interfaces to education. Among them, \textbf{E1} and \textbf{E5} further participated in two case studies (Section~\ref{sec: casestudy}).\revise{In addition, we quantitatively evaluated the proposed attention quantification model and concept relationship extraction method to verify their performance, and conducted a sensitivity analysis to confirm the robustness of our content-derived metrics.}

\subsection{Case Study} \label{sec: casestudy}
This section presents two case studies conducted by \textbf{E1} and \textbf{E5} to demonstrate the effectiveness of \techName{}.

\textbf{Case 1. Identifying Causes of Low Attention: Insufficient or Excessive.} \textbf{E1} expressed particular interest in the course on \textit{Convolutional Neural Networks} and aimed to leverage \techName{} to identify the causes of attention decline. Upon launching \techName{}, he first explored the \textit{Concept Module} (Fig.~\ref{fig:case1}\casepic{A}) to obtain an overview of course structure and attention distribution across concept nodes.
He focused on important concepts, characterized by a larger dark-gray inner ring~\grayring{} radius and a higher angular proportion of the arc indicator~\halfring{} beneath the node label. Among them, he quickly identified \textit{Sliding Computation of Convolutional Kernels} (Fig.~\ref{fig:case1}\casepic{A1}). Its blue~\bluesquare{}  segmented ring shows both a large angular span and height, suggesting students experienced prolonged attention deviation while learning this concept.

\textit{\textbf{Information Density.}} 
\revise{In Fig.~\ref{fig:case1}\casepic{A1}, \textbf{E1} observed that the dark-green with white stripe progress bar~\greendensity{} of this concept node occupied a noticeably small angle, preliminarily suggesting that low student attention might stem from insufficient visual information density.} He clicked on the concept to focus on its corresponding segment in the \textit{Group Module} (Fig.~\ref{fig:case1}\casepic{B}) for further investigation.In Fig.~\ref{fig:case1}\casepic{B2}, the barcodes with blue~\bluesquare{} outlines in the attention track were notably long, indicating students were in a low attention state for most of the time. When examining the audiovisual tracks, he observed that the purple polyline~\purplepolyline{} representing AAC beneath the barcodes remained relatively stable, whereas the dark green polyline~\greenpolyline{} representing VAC in the visual track above (dashed box in Fig.~\ref{fig:case1}\casepic{B2}) exhibited a noticeable dip, suggesting this as a primary cause of the attention decline. Moreover, the two dark green density rectangles above the polyline had nearly vanished~\framefive{}, leaving only two downward triangles, indicating a significant drop in visual information density. 

\revise{To verify the actual visual information density during this period, \textbf{E1} selected the four dark-green density rectangles with white stripes~\stripedbox{} in the 7:58–10:52 window. The corresponding original video clips were then pinned to the video annotation panel and played in sync (Fig.~\ref{fig:case1}\casepic{D}). The four clips showed successive presentation stages of the same slide. Initially, the slide contained explanatory text and a CNN flowchart, resulting in a normal level of visual information density. After this content had been introduced, much of it disappeared during the transition, and the slide subsequently remained nearly blank until additional figures and text were progressively presented. This near-blank state corresponded to the nearly vanished staircase line beneath the dashed box in Fig.~\ref{fig:case1}\casepic{B2}. These checks confirmed that the visual information density decreased substantially during this period. Combined with the previously identified decline in learner attention, \textbf{E1} suggested retaining the existing visual content during the transition and shortening the time spent on the near-blank slide.}
\textbf{E1} therefore hypothesized that insufficient visual information density contributed to the decline in student attention. 
\revise{The experts inspected the source video throughout the analysis, most often at periods of abnormal information density, following the same verify and annotate routine, so we omit these repeated steps below.}

To seek further evidence, \textbf{E1} continued exploring the audiovisual tracks. He identified a similar pattern in the auditory track (Fig.~\ref{fig:case1}\casepic{B4}), where downward triangles accompanied by long straight lines and vanishing purple density rectangles~\framesix{} indicated a rapid decline in auditory density. By examining the purple polyline~\purplepolyline{} fluctuations, he hypothesized that the extended blue-outlined~\bluesquare{} barcodes (Fig.~\ref{fig:case1}\casepic{B4}) were caused by the drop in auditory density. However, he soon noticed a large area of blue-outlined~\bluesquare{} barcodes in Fig.~\ref{fig:case1}\casepic{B5}. Counterintuitively, ten long rectangles appeared in the audiovisual tracks (dashed box in Fig.~\ref{fig:case1}\casepic{B5}), indicating substantially high audiovisual density. \textbf{E1} commented, ``\textit{Excessively high information volume may exceed students' cognitive load capacity.}'' Indeed, he found similar patterns in the dashed boxes of Fig.~\ref{fig:case1}\casepic{B1} and \casepic{B6}. Meanwhile, he noticed that the density levels triggering attention changes were not consistent across segments. Based on these clues, \textbf{E1} speculated that students' cognitive load exhibited a pattern of being low at the beginning and end but high in the middle, and that both excessively low and excessively high information density could lead to a decline in student attention.

\textit{\textbf{Concept Relationships.}} In Fig.~\ref{fig:case1}\casepic{B3}, \textbf{E1} noticed an unanalyzed segment of blue-outlined~\bluesquare{} barcodes. He examined the rectangles in the audiovisual tracks above and below but found no significant changes in their lengths. Turning to Fig.~\ref{fig:case1}\casepic{A2}, he spotted a dashed line~\bluedashedline{} representing a derivation relationship, linking the current concept with \textit{Sliding Computation of Convolutional Kernels}. \textbf{E1} hypothesized that this prerequisite relationship might have hindered students' comprehension, and sought stronger evidence through the \textit{Individual Module} (Fig.~\ref{fig:case1}\casepic{C}). He dragged the time locator bar and adjusted the temporal resolution to 1s to inspect each student's attention changes at finer granularity. He then clicked on a streamline~\blueflow{} within the blue region in the focused segment of \textit{Sliding Computation} (Fig.~\ref{fig:case1}\casepic{C1}), where a hovering black label indicated it belonged to \textit{Student~5}. A highlighted blue streamline~\blueflow{} appeared in the thumbnail at the lower-right corner of both Fig.~\ref{fig:case1}\casepic{C1} and \casepic{C2}, indicating this student maintained low attention while studying both concepts. \textbf{E1} continued clicking on more streamlines within the blue~\blueregion{} region, and the majority exhibited a similar pattern. Based on these findings, \textbf{E1} concluded that mind-wandering during prerequisite concepts is also a contributing factor to low attention.

\textit{\textbf{Root Cause.}} While exploring the \textit{Individual Module}, \textbf{E1} progressively zoomed into Fig.~\ref{fig:case1}\casepic{C2} and discovered that streamline overlap differed notably across regions: widest in the pink~\pinkregion{} region and narrowest in the blue~\blueregion{}, indicating that higher attention corresponded to higher SAC. He then separately selected streamlines from the blue~\blueregion{}, yellow~\yellowregion{}, and pink~\pinkregion{} regions (Fig.~\ref{fig:case1}\casepic{E4}--\casepic{E6}), producing the results in Fig.~\ref{fig:case1}\casepic{E1}--\casepic{E3}: when one channel's contribution was similar, greater contribution from the other corresponded to greater SAC and higher attention levels. The same pattern was confirmed in Fig.~\ref{fig:case1}\casepic{C1}. In fact, variations in audiovisual information density ultimately influenced attention by affecting SAC.

In summary, \textbf{E1} concluded through Case 1 that: (1) attention level is positively correlated with SAC; (2) both excessively low and high information density can reduce attention contribution; and (3) mind-wandering during prerequisite concepts can cause attention fluctuations in downstream concepts.

\begin{figure}[!t]
  \centering
  \setlength{\belowcaptionskip}{-0.6cm}
  \includegraphics[width=\columnwidth]{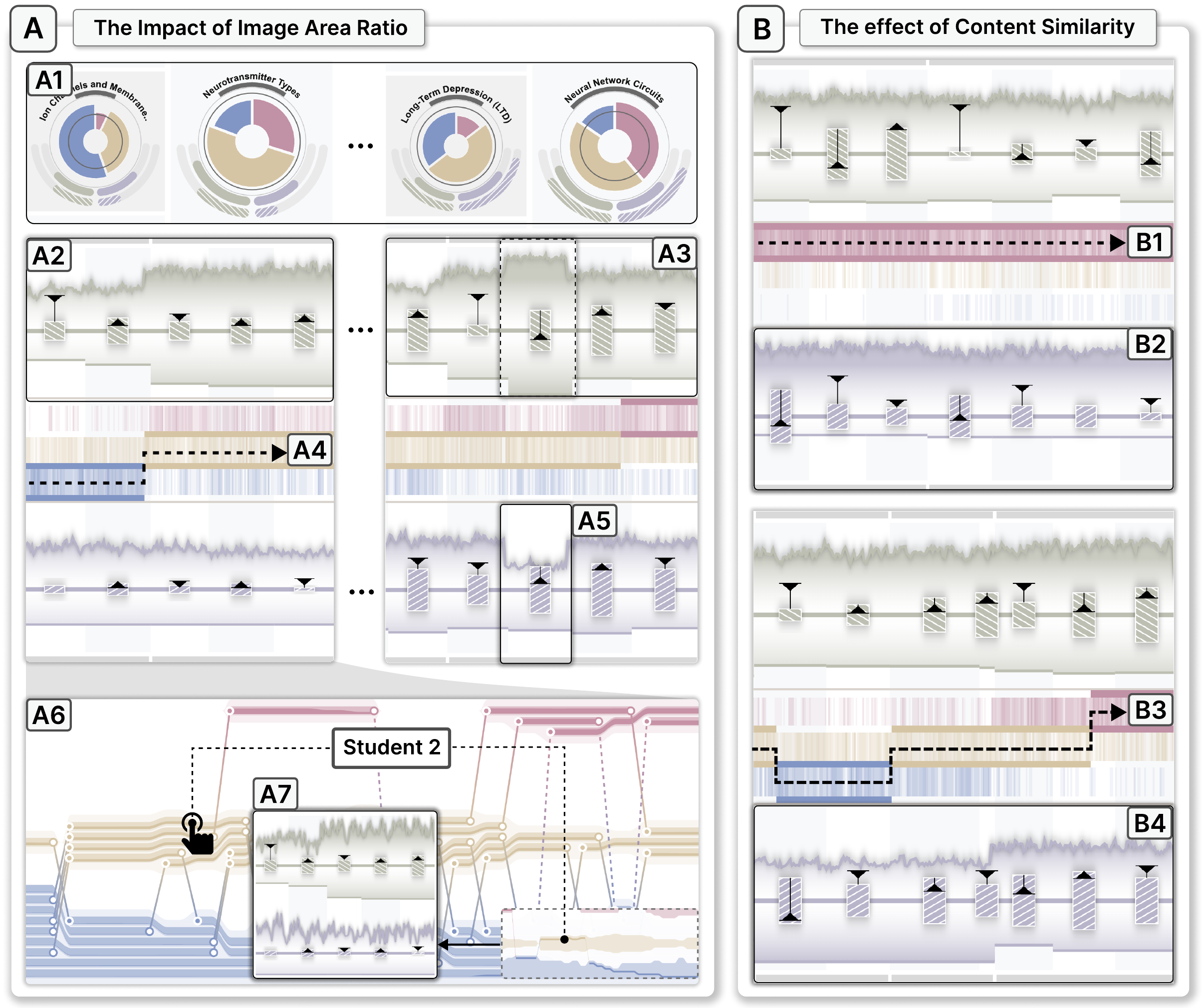}
  \caption{A teacher uses \techName{} to explore strategies behind high attention. (A) displays attention distribution across concepts (A1). The teacher discovers the impact of image area ratio on attention through the visual track (A2--A5), and verifies these findings by comparing individual students' audiovisual attention contributions in the \textit{Individual Module} (A6--A7). (B) The auditory track reveals the effect of narration-slide redundancy on attention (B1--B4).}
  \label{fig:case2}
\end{figure}


\textbf{Case 2. Exploring Strategies for High Attention: The Slide Effect.} With a background in neuroscience, \textbf{E5} selected the course on \textit{Neurons and Synapses}. Unlike Case~1, by examining the \textit{Concept Module} (Fig.~\ref{fig:case2}\casepic{A}), he found that most students maintained high attention. He therefore decided to use the \textit{Group Module} directly as the entry point to explore the strategies behind high attention. To better observe audiovisual attention contribution changes, \textbf{E5} repositioned the information density rectangles onto the audiovisual tracks' x-axis.

\textit{\textbf{Image Area Ratio.}} In the selected region of Fig.~\ref{fig:case2}\casepic{A2}, \textbf{E5} noticed an interesting pattern: the green polyline~\greenpolyline{} and the staircase line~\greenstaircase{} below it formed a near-symmetric shape, while the green rectangles~\stripedbox{} on the x-axis showed no significant change in length. This pattern was even more pronounced in Fig.~\ref{fig:case2}\casepic{A3}. He also observed that in the attention track below (Fig.~\ref{fig:case2}\casepic{A4}), the outline color transitioned from blue~\bluesquare{} to yellow~\yellowsquare{}, leading him to hypothesize that the deepening of the dark green staircase depth, i.e., the increase in image area ratio, can effectively improve student attention.
To validate this, he brushed the corresponding time range and navigated to the \textit{Individual Module }(Fig.~\ref{fig:case2}\casepic{A6}). Similar to \textbf{E1}'s approach, he clicked on a streamline labeled \textit{Student~2} in the yellow~\yellowregion{} region. The audiovisual tracks transformed into Fig.~\ref{fig:case2}\casepic{A7}, where the green polyline~\greenpolyline{} and staircase line~\greenstaircase{} still maintained their near-symmetric shape, indicating \textit{Student~2}'s VAC trend was consistent with the group-level pattern. By clicking on more streamlines, he confirmed the majority exhibited the same behavior.         \revise{\textbf{E5} was inclined to think that increasing the image area ratio in slides can effectively enhance students' VAC.}

However, \textbf{E5} noticed that the attention track below the dashed box in Fig.~\ref{fig:case2}\casepic{A3} showed no significant improvement in attention, even though the green staircase~\greenstaircase{} depth was already substantial. Intrigued, he observed that in Fig.~\ref{fig:case2}\casepic{A5}, the purple polyline~\purplepolyline{} exhibited a noticeable dip, indicating a rapid decline in AAC. Meanwhile, the purple density rectangles~\purplestripedbox{} showed no significant changes, leading him to suspect that slide design might affect not only visual but also auditory attention. In fact, dominant visual information can ``capture'' attentional resources at the expense of auditory processing, a phenomenon known as inattentional deafness.

\textit{\textbf{Content Similarity.}} \textbf{E5} then shifted focus to the auditory track. Comparing the solid boxes in Fig.~\ref{fig:case2}\casepic{B2} and Fig.~\ref{fig:case2}\casepic{B4}, he noticed that, unlike the visual track, the lower the purple staircase line~\purplestaircase{}, the higher the purple polyline~\purplepolyline{} representing AAC. He also found that green polyline~\greenpolyline{} heights in both segments were similar, indicating comparable VAC. Meanwhile, the stark contrast between the pink~\pinksquare{} thick-outlined barcodes in Fig.~\ref{fig:case2}\casepic{B1} and the gradually transitioning tri-colored barcodes in Fig.~\ref{fig:case2}\casepic{B3} further corroborated this. Based on \textbf{E1}'s conclusion in Case~1, \textbf{E5} inferred that the barcode outline differences stemmed from changes in AAC, and concluded that the similarity between narration and slide text may be negatively correlated with AAC. He verified this by clicking on individual streamlines in the \textit{Individual Module}.

After analyzing the \textit{Neurons and Synapses} using \techName{}, \textbf{E5} deduced that: (1) appropriately increasing the image area ratio in slides can enhance student attention, as images are more effective at capturing attention; and (2) information redundancy should be avoided in instructional design, meaning narration and slide text should minimize overlap, allowing the eyes and ears to serve complementary roles.

\begin{table*}[t]
  \centering
  \setlength{\belowcaptionskip}{-0.4cm}
  \caption{The evaluation results of our proposed model on four datasets over five runs.}
  \label{tab:model_eval_results}
  \scriptsize
  \setlength{\tabcolsep}{3.0pt}
  \renewcommand{\arraystretch}{1.12}
  \resizebox{\textwidth}{!}{
  \begin{tabular}{lcccccc}
    \toprule
    Dataset 
    & \textbf{Our Model} 
    & Our Model\textsubscript{w/o EEG}
    & Our Model\textsubscript{w/o fNIRS}
    & SVM(2022)~\cite{apicella2022eeg}
    & TASNet+DRL(2023)~\cite{zhang2023unsupervised}
    & DDQN(2025)~\cite{rehman2025measuring} \\
    \midrule
    Convolutional Neural Networks 
    & \textbf{96.94 $\pm$ 0.16} 
    & 87.18 $\pm$ 0.47 
    & 77.19 $\pm$ 0.36 
    & 71.27 $\pm$ 2.63 
    & 88.60 $\pm$ 0.20 
    & \underline{90.11 $\pm$ 0.78} \\

    Matrix Factorization 
    & \textbf{97.76 $\pm$ 0.12} 
    & 87.93 $\pm$ 0.48 
    & 78.40 $\pm$ 0.61 
    & 72.49 $\pm$ 1.30 
    & 89.40 $\pm$ 0.11 
    & \underline{94.55 $\pm$ 0.41} \\

    Neurons and Synapses 
    & \textbf{97.16 $\pm$ 0.86} 
    & 85.73 $\pm$ 0.52 
    & 79.74 $\pm$ 0.83 
    & 71.75 $\pm$ 0.92 
    & 89.59 $\pm$ 0.19 
    & \underline{92.16 $\pm$ 1.21} \\

    Driving Dataset~\cite{aci2019distinguishing} 
    & \textbf{98.85 $\pm$ 0.88} 
    & -- 
    & \textbf{98.85 $\pm$ 0.88} 
    & 76.82 $\pm$ 3.43 
    & 81.70 $\pm$ 1.42 
    & \underline{97.25 $\pm$ 2.37} \\
    \bottomrule
  \end{tabular}
  }
  \vspace{-0.5cm}
\end{table*}

\subsection{Model Evaluation}
We conducted modality ablation and comparative experiments to validate the effectiveness of multimodal brain data and demonstrate superiority over existing methods. \revise{Computational efficiency and hyperparameter evaluations are provided in Supplementary Materials.}

\textbf{Experiment Setup.}
We evaluated the model under \textit{EEG-only}, \textit{fNIRS-only}, and \textit{hybrid} configurations, and compared against SVM~\cite{apicella2022eeg}, TASNET+DRL~\cite{zhang2023unsupervised}, and DDQN~\cite{rehman2025measuring}, using brain data from three course sessions (Section~\ref{sec:Brain Dataset}) with ground-truth labels constructed following the protocol therein, and a driving EEG benchmark~\cite{aci2019distinguishing} with its native labels for generalizability evaluation. We report accuracy (ACC) with standard deviations over five runs.

\textbf{Results.}
Table~\ref{tab:model_eval_results} reports performance averaged over five runs. The model achieved over \textbf{95\%} accuracy on all three instructional video datasets, fully meeting expert expectations. The ablation study confirms that the multimodal configuration outperforms single-modality variants, while comparative experiments show it surpasses all baselines. The model also delivers competitive performance on the driving dataset, highlighting robust generalization. Since this dataset contains only EEG signals, Our Model is equivalent to Our Model\textsubscript{w/o fNIRS}, so the two columns report identical results. Our Model\textsubscript{w/o EEG} is not applicable and marked as N/A.
Overall, these results verify the robustness and effectiveness of our model across datasets and input settings.

\begin{table}[t]
\centering
\caption{Sensitivity analysis of the information density.}
\label{tab:sensitivity}
\setlength{\tabcolsep}{4pt}
\begin{tabular}{@{}llccc@{}}
\toprule
Metric (default) & Alternative & $\rho$ & Jaccard & TA \\
\midrule
\multirow{7}{*}{Visual info.\ density}
 & Edge $+$ entropy              & 0.93 & 0.78 & 0.87 \\
 & Edge $+$ type--token          & 0.89 & 0.72 & 0.85 \\
 & Clutter $+$ IDF               & 0.82 & 0.69 & 0.77 \\
 & Clutter $+$ entropy           & 0.84 & 0.71 & 0.82 \\
 & Clutter $+$ type--token       & 0.81 & 0.63 & 0.74 \\
 & Equal weights                 & 0.86 & 0.75 & 0.83 \\
 & Normalized area-ratio weights & 0.95 & 0.78 & 0.92 \\
\midrule
\multirow{3}{*}{Auditory info.\ density}
 & Speech rate           & 0.87 & 0.62 & 0.73 \\
 & Type--token ratio     & 0.75 & 0.68 & 0.80 \\
 & Total self-information & 0.86 & 0.74 & 0.79 \\
\bottomrule
\end{tabular}
\end{table}

\revise{\subsection{Sensitivity Analysis}
We assess the robustness of our two content-derived metrics, the visual and auditory information density, on 20 randomly sampled slide-based videos from Coursera.\footnote{\url{https://www.coursera.org}} At each step of their integration, we substitute literature-backed alternatives and check whether the resulting patterns remain stable.

\textbf{Experiment Setup.}
For the \textbf{\textit{visual information density}}, we vary how the image and text channels are computed and combined. For image complexity we add subband-entropy clutter~\cite{rosenholtz2007measuring}, and for textual information token-distribution entropy~\cite{shannon1948mathematical} and type--token ratio~\cite{mccarthy2010mtld}, giving five image--text combinations beyond our default. We also replace the area-ratio weights with equal weights and renormalized area ratios. For the \textbf{\textit{auditory information density}}, we substitute word-distribution entropy with speech rate~\cite{coupe2019different}, type--token ratio~\cite{mccarthy2010mtld}, and total self-information~\cite{shannon1948mathematical}.

We assess stability along three dimensions: global ranking agreement (Spearman's $\rho$)~\cite{spearman1961proof}, top-$k$ Jaccard overlap on the most salient segments ($k$=top $20\%$ of one-second intervals)~\cite{jaccard1912distribution}, and trend agreement(TA) on the direction of change between consecutive segments(relative change below $\varepsilon=5\%$ as unchanged)~\cite{kendall1938new}. For the top-$k$ overlap, we treat each one-second interval as a segment and set $k$ to the top $20\%$ most information-dense segments of each video. For trend agreement, a transition is labeled as approximately unchanged when its relative change is below $\varepsilon=5\%$. All measures are computed per video and averaged over the 20 videos.

\textbf{Results.}
Table~\ref{tab:sensitivity} reports the results, averaged over the 20 videos. Across all alternatives, the recomputed densities remain strongly aligned with our default. For the visual information density, the five alternatives attain an average Spearman rank correlation of $0.87$, a top-$k$ Jaccard overlap of $0.72$, and a trend agreement of $0.83$. For the auditory information density, the three alternatives attain $0.83$, $0.68$, and $0.77$, respectively. These results show that the ordering of segments by information density, the most information-dense segments, and the temporal trends are jointly preserved across the alternative operationalizations.
}

\subsection{Concept Relationship Extraction}\label{sec:Extraction}
Because we replaced the LLM used for concept relationship extraction in ConceptThread~\cite{zhou2024conceptthread} from GPT-3.5 with Gemini-3-pro-preview, it was necessary to re-evaluate the updated extraction performance.
However, concept relationships in instructional videos lack widely accepted ground truth, making direct evaluation of generated results difficult~\cite{zhou2024conceptthread}.
We therefore invited two experts (\textbf{E1} and \textbf{E2}) to rate over 50 sets of concept relationships extracted by the system from 10 instructional videos familiar to them on a 1--7 Likert scale (1 = ``totally unacceptable'', 7 = ``very good''). The results indicate that the updated method performs well for concept relationship extraction (\textbf{E1}: 5.5, \textbf{E2}: 6.3).

\subsection{Expert Interview}\label{sec: interview}
\textbf{Procedure.} To validate the usability and effectiveness of \techName{}, we conducted semi-structured interviews with six domain experts (\textbf{E1--E6}), each lasting approximately 70 minutes. All experts had experience in online teaching or learning analytics research. We first briefly introduced the research background and provided a tutorial to demonstrate the visual design, interactions, and analytic workflow of \techName{} (30 minutes). Afterward, experts were allowed to freely explore the system using data collected from three real online courses with varying subjects and durations (Section~\ref{sec:Brain Dataset}) (20 minutes). Given that some experts were unfamiliar with visual analytics, we adopted the co-discovery method to help them get started. Subsequently, we conducted individual interviews with each expert (20 minutes), focusing on three aspects: the effectiveness of the visual design in supporting attention analysis, the practicality of the analytic workflow in real teaching scenarios, and suggestions for further improvement.

\textbf{System Design.}
All experts (\textbf{E1--E6}) regarded \techName{} as a useful tool for VBL optimization. With a brief tutorial, they quickly grasped our visual design and analytic workflow.
\textbf{E1--E3} praised the \textit{Concept Module} for helping teachers quickly review course structure and screen concepts for revision, avoiding inefficient frame-by-frame review.
\revise{\textbf{E5} added, ``\textit{When the \textit{Concept Module} does not reveal clear differences among concepts, the \textit{Group Module} provides an alternative analytic entry point.}''} \textbf{E1} noted that conventional video revision requires substantial time and experience, often involving multiple rounds of trial and error. \revise{\textbf{E2} stated, ``\textit{How to configure images and text in slides, and how to write narration to complement the slides, these criteria are often ambiguous, and sometimes even experience can be misleading.}''}
Through the \textit{Group Module}, he could quickly pinpoint audiovisual design weaknesses and draft revision plans.
\textbf{E4--E6} gave highly positive feedback, particularly appreciating the temporal linkage across modules.
The \textit{Individual Module} was similarly well received. Experts found it not only presents each student's attention trajectory, but also allows teachers to verify group-level hypotheses through dynamic temporal resolution.
\textbf{E4} noted that \techName{} offered new research insights, ``\textit{Visualization opens a new perspective for our research.}'' 
Exploring the \textit{Individual Module}, he observed some students consistently maintained high VAC but low AAC, suggesting different audiovisual preference types may exist. He believed balancing the majority and those with pronounced preferences warrants further investigation, and expressed interest in incorporating more audiovisual features into attention modeling and leveraging visualization to better communicate findings.

\textbf{Usability and Suggestions.}
All experts agreed that \techName{}'s data processing pipeline and analytic workflow align with established practices in neuroscience and education.
\revise{\textbf{E6} commented, ``\textit{DDQN captures subtle attention transition patterns through reinforcement learning, and EEG--fNIRS fusion combines high temporal sensitivity with strong robustness.}'' In his view, these substantially improve usability in real-world settings.} Although experts considered all modules easy to use, they offered several suggestions. \revise{\textbf{E5} noted that the \textit{Group Module} appears slightly complex initially, but its value becomes apparent after familiarization. He remarked, ``\textit{This design is highly informative for analyzing attention alongside audiovisual stimulus variation. It would be better if users could understand it without any tutorial.}''}
He also suggested supporting import of multi-round revision data for comparative analysis.
\textbf{E4} suggested annotating teaching strategies (e.g., metaphor and exemplification) for each concept, as they are another key factor influencing attention.

\section{Discussion}
In this section, we discuss the lessons learned during the development and evaluation of \techName{}, and outlines our limitations.

\textbf{Improvements on the Current Workflow.}
\techName{} introduces several improvements for VBL optimization.
\revise{Unlike existing approaches that rely on self-reports such as questionnaires or behavioral signals such as eye-tracking, \techName{} leverages brain signals to reflect fluctuations in learners' attention, enabling multi-level quantitative analysis from course content structure to individual attention. Going a step further, it reveals deeper associations between instructional design and student attention.}
Moreover, \techName{} substantially reduces the time required for course optimization. Teachers often struggle to extract meaningful patterns from high-dimensional temporal attention data, whereas \techName{} enables exploration and comparison of attention dynamics from multiple perspectives at varying granularities.
Additionally, \techName{} provides a new perspective for attention-related research, leveraging real-world data and novel visualization designs to present the effects of audiovisual stimuli on attention in greater depth.

\textbf{Learning Curve.}
Our expert interview suggested that reducing learning costs is critical. All experts preferred intuitive solutions to simplify the transition from data reading to pattern interpretation. Accordingly, we balanced information richness and usability in the visual design. For example, the \textit{Concept Module} serves as the analytical entry point because it matches users' mental model of course structure, while avoiding excessive glyph-level embellishment. The \textit{Group Module} and \textit{Individual Module} were inspired by line charts and streamgraphs, effectively shortening the learning curve. The overall workflow follows the visualization mantra of Overview First and Details on Demand~\cite{shneiderman2003eyes}, helping users easily identify patterns of student attention dynamics. After approximately 15 minutes of tutorial, experts were able to understand each module and the interactions within \techName{}.

\textbf{Generalizability.}
As an initial attempt to visually interpret student attention, \techName{} currently focuses on typical VBL scenarios, but its core framework holds broader application potential.
First, \techName{} can also help students understand course knowledge structure, identify attention weaknesses, and support self-monitoring and adaptive learning strategy adjustment.
Second, the framework can extend beyond education. In film and video production, it could help creators analyze audience attention and emotional arousal patterns to refine narrative pacing and audiovisual orchestration. In clinical settings, it may support ADHD assessment~\cite{hager2021biomarker} and pre-seizure signal pattern identification in epilepsy~\cite{pellinen2024improving}.
Overall, although \techName{} is currently grounded in VBL optimization, its design rationale and analytical framework show strong cross-scenario transfer potential.

\textbf{Limitations.}
\techName{} is not without limitations.
Scalability may become challenging as the number of concepts increases with video duration, and the \textit{Individual Module} can become visually crowded with larger student samples. Our path grouping and merging strategy, together with scrollbar interaction, partially alleviates this.
Although \techName{} links attention dynamics to course content, it currently only supports problem identification without generating specific optimization recommendations.
\revise{Experts also noted that the visual and auditory information density are computed from the presented instructional materials, and therefore reflect the information delivered by the slides and narration rather than the specific content that each learner actually processes. Finally, our audiovisual decomposition does not directly measure top-down attention, but rather provides surrogate indicators biased toward top-down attentional allocation.}

\section{Conclusion and Future Work}
We presented \techName{}, an interactive visual analytics system designed to help teachers explore patterns of student attention variation in VBL. Specifically, we first constructed a deep reinforcement learning-based attention quantification model using multimodal brain data, and then designed a top-down visual analytics workflow comprising the \textit{Concept Module}, \textit{Group Module}, and \textit{Individual Module}, enabling teachers to analyze from course content and audiovisual design to individual attention trajectories. We conducted quantitative model evaluation, two case studies, and in-depth interviews with six domain experts to validate the effectiveness and usability of \techName{}. The results demonstrate that \techName{} can effectively support teachers in identifying attention variation patterns and provide valuable analytical insights for course optimization.

In future work, we plan to develop methods that automatically screen and aggregate highly representative student data to optimize the data scale. We will add more quantifiable indicators that may influence attention, such as teaching strategy types. \revise{In addition, we will refine the information density metrics at the learner level to obtain learner-specific, real-time estimates of information intake, for example by adopting eye tracking for the visual channel and cortical speech tracking for the auditory channel~\cite{geirnaert2021}.}\revise{Finally, we plan to use \techName{} in more naturalistic settings to study how learners' study behaviors, such as note-taking, reshape their attention patterns and building on these patterns to help teachers infer a learner's current study behavior.}

\bibliographystyle{IEEEtran}
\bibliography{template}

@article{sablic2021video,
  title={Video-based learning (VBL)—past, present and future: An overview of the research published from 2008 to 2019},
  author={Sabli{\'c}, Marija and Mirosavljevi{\'c}, Ana and {\v{S}}kugor, Alma},
  journal={Technology, Knowledge and Learning},
  volume={26},
  number={4},
  pages={1061--1077},
  year={2021},
  publisher={Springer},
  doi       = {10/grb6kk}
}

@inproceedings{shi2015vismooc,
  title={VisMOOC: Visualizing video clickstream data from massive open online courses},
  author={Shi, Conglei and Fu, Siwei and Chen, Qing and Qu, Huamin},
  booktitle={2015 IEEE Pacific visualization symposium (PacificVis)},
  pages={159--166},
  year={2015},
  organization={IEEE},
  doi       = {10/hbt7sj}
}

@article{seaton2014does,
  title={Who does what in a massive open online course?},
  author={Seaton, Daniel T and Bergner, Yoav and Chuang, Isaac and Mitros, Piotr and Pritchard, David E},
  journal={Communications of the ACM},
  volume={57},
  number={4},
  pages={58--65},
  year={2014},
  publisher={ACM New York, NY, USA}
}

@article{o2015use,
  title={The use of flipped classrooms in higher education: A scoping review},
  author={O'Flaherty, Jacqueline and Phillips, Craig},
  journal={The internet and higher education},
  volume={25},
  pages={85--95},
  year={2015},
  publisher={Elsevier},
  doi       = {10/f665sg}
}

@article{wieling2010impact,
  title={The impact of online video lecture recordings and automated feedback on student performance},
  author={Wieling, Martijn B and Hofman, Wiecher Hermen Adriaan},
  journal={Computers \& Education},
  volume={54},
  number={4},
  pages={992--998},
  year={2010},
  publisher={Elsevier},
  doi       = {10/czdwcr}
}

@article{zhou2024conceptthread,
  title={Conceptthread: visualizing threaded concepts in MOOC videos},
  author={Zhou, Zhiguang and Ye, Li and Cai, Lihong and Wang, Lei and Wang, Yigang and Wang, Yongheng and Chen, Wei and Wang, Yong},
  journal={IEEE transactions on visualization and computer graphics},
  volume={31},
  number={2},
  pages={1354--1370},
  year={2024},
  publisher={IEEE},
  doi       = {10/hbt7sk}
}

@article{weston1995model,
  title={A model for understanding formative evaluation in instructional design},
  author={Weston, Cynthia and McAlpine, Lynn and Bordonaro, Tino},
  journal={Educational technology research and development},
  volume={43},
  number={3},
  pages={29--48},
  year={1995},
  publisher={Springer},
  doi       = {10/dtf5pt}
}

@article{norman2017twelve,
  title={Twelve tips for reducing production time and increasing long-term usability of instructional video},
  author={Norman, Marie K},
  journal={Medical teacher},
  volume={39},
  number={8},
  pages={808--812},
  year={2017},
  publisher={Taylor \& Francis},
  doi       = {10/ggmhdg}
}

@inproceedings{von2019iterative,
  title={Iterative course design in MOOCs: Evaluating a protoMOOC},
  author={von Schmieden, Karen and Mayer, Lena and Taheri, Mana and Meinel, Christoph},
  booktitle={Proceedings of the Design Society: International Conference on Engineering Design},
  volume={1},
  number={1},
  pages={539--548},
  year={2019},
  organization={Cambridge University Press},
  doi       = {10/hbt7sn}
}

@article{fredricks2004school,
  title={School engagement: Potential of the concept, state of the evidence},
  author={Fredricks, Jennifer A and Blumenfeld, Phyllis C and Paris, Alison H},
  journal={Review of educational research},
  volume={74},
  number={1},
  pages={59--109},
  year={2004},
  publisher={Sage Publications Sage CA: Thousand Oaks, CA},
  doi       = {10/btdzg6}
}

@article{brame2017effective,
  title={Effective educational videos: Principles and guidelines for maximizing student learning from video content},
  author={Brame, Cynthia J},
  journal={CBE—Life Sciences Education},
  year={2017},
  publisher={American Society for Cell Biology},
  doi       = {10/gf5nz8},
  volume    = {15},
  number    = {4}
}

@article{chavan2022tcherly,
  title={Tcherly: A Teacher-Facing Dashboard for Online Video Lectures.},
  author={Chavan, Pankaj and Mitra, Ritayan},
  journal={Journal of Learning Analytics},
  volume={9},
  number={3},
  pages={125--151},
  year={2022},
  publisher={ERIC},
  doi       = {10/nxsx}
}

@article{keegan1995distance,
  title={Distance education technology for the new millennium: Compressed video teaching},
  author={Keegan, Desmond},
  year={1995},
  journal   = {ZIFF Papiere},
  volume    = {101},
  publisher = {FernUniversit{"{a}}t Hagen}
}

@article{aggarwal2021preliminary,
  title={A preliminary investigation for assessing attention levels for massive online open courses learning environment using eeg signals: An experimental study},
  author={Aggarwal, Swati and Lamba, Mohit and Verma, Kandarp and Khuttan, Siddharth and Gautam, Hitesh},
  journal={Human Behavior and emerging technologies},
  volume={3},
  number={5},
  pages={933--941},
  year={2021},
  publisher={Wiley Online Library},
  doi       = {10/gnsnkm}
}

@inproceedings{duval2011attention,
  title={Attention please! Learning analytics for visualization and recommendation},
  author={Duval, Erik},
  booktitle={Proceedings of the 1st international conference on learning analytics and knowledge},
  pages={9--17},
  year={2011},
  doi       = {10/fxvbww}
}

@inproceedings{navarro2024vaad,
  title={VAAD: Visual attention analysis dashboard applied to e-learning},
  author={Navarro, Miriam and Becerra, {\'A}lvaro and Daza, Roberto and Cobos, Ruth and Morales, Aythami and Fierrez, Julian},
  booktitle={2024 International Symposium on Computers in Education (SIIE)},
  pages={1--6},
  year={2024},
  organization={IEEE},
  doi       = {10/hbt7ss}
}

@article{sabuncuoglu2023developing,
  title={Developing a multimodal classroom engagement analysis dashboard for higher-education},
  author={Sabuncuoglu, Alpay and Sezgin, T Metin},
  journal={Proceedings of the ACM on Human-Computer Interaction},
  volume={7},
  number={EICS},
  pages={1--23},
  year={2023},
  publisher={ACM New York, NY, USA},
  doi       = {10/hbt7st}
}

@inproceedings{sharma2016gaze,
  title={A gaze-based learning analytics model: in-video visual feedback to improve learner's attention in MOOCs},
  author={Sharma, Kshitij and Alavi, Hamed S and Jermann, Patrick and Dillenbourg, Pierre},
  booktitle={Proceedings of the sixth international conference on learning analytics \& knowledge},
  pages={417--421},
  year={2016},
  doi       = {10/gk5vzk}
}

@article{deng2023review,
  title={A review of eye tracking research on video-based learning},
  author={Deng, Ruiqi and Gao, Yifan},
  journal={Education and Information Technologies},
  volume={28},
  number={6},
  pages={7671--7702},
  year={2023},
  publisher={Springer},
  doi       = {10/gr5dj4}
}

@article{xie2025eeg,
  title={EEG analysis of brain dynamics in a simulated multi-task and multi-stage learning environment},
  author={Xie, Hui and Jia, Chunli and Luo, Yanxia and He, Jiangshan and Dong, Zexiao and Liang, Dan and Ren, Ziqi and Jiang, Mingzhe and Gao, Xinbo and Chen, Xueli},
  journal={npj Science of Learning},
  volume={10},
  number={1},
  pages={84},
  year={2025},
  publisher={Nature Publishing Group UK London},
  doi       = {10/hbt7sv}
}

@article{shaw2023attention,
  title={Attention classification and lecture video recommendation based on captured EEG signal in flipped learning pedagogy},
  author={Shaw, Rabi and Patra, Bidyut Kr and Pradhan, Animesh and Mishra, Swayam Purna},
  journal={International Journal of Human--Computer Interaction},
  volume={39},
  number={15},
  pages={3057--3070},
  year={2023},
  publisher={Taylor \& Francis},
  doi       = {10/hbt7sw}
}

@article{xu2022eeg,
  title={EEG data quality in real-world settings: Examining neural correlates of attention in school-aged children},
  author={Xu, Keye and Torgrimson, Sarah Jo and Torres, Remi and Lenartowicz, Agatha and Grammer, Jennie K},
  journal={Mind, brain, and Education},
  volume={16},
  number={3},
  pages={221--227},
  year={2022},
  publisher={Wiley Online Library},
  doi       = {10/gpsxpw}
}

@article{andrienko2012visual,
  title={Visual analytics methodology for eye movement studies},
  author={Andrienko, Gennady and Andrienko, Natalia and Burch, Michael and Weiskopf, Daniel},
  journal={IEEE transactions on Visualization and Computer Graphics},
  volume={18},
  number={12},
  pages={2889--2898},
  year={2012},
  publisher={IEEE},
  doi       = {10/bcnq}
}

@article{henrie2015measuring,
  title={Measuring student engagement in technology-mediated learning: A review},
  author={Henrie, Curtis R and Halverson, Lisa R and Graham, Charles R},
  journal={Computers \& Education},
  volume={90},
  pages={36--53},
  year={2015},
  publisher={Elsevier},
  doi       = {10/gfxcmq}
}

@article{giannakos2019multimodal,
  title={Multimodal data as a means to understand the learning experience},
  author={Giannakos, Michail N and Sharma, Kshitij and Pappas, Ilias O and Kostakos, Vassilis and Velloso, Eduardo},
  journal={International Journal of Information Management},
  volume={48},
  pages={108--119},
  year={2019},
  publisher={Elsevier},
  doi       = {10/gg93wt}
}

@inproceedings{rudovic2019personalized,
  title={Personalized estimation of engagement from videos using active learning with deep reinforcement learning},
  author={Rudovic, Ognjen and Park, Hae Won and Busche, John and Schuller, Bj{\"o}rn and Breazeal, Cynthia and Picard, Rosalind W},
  booktitle={2019 IEEE/CVF Conference on Computer Vision and Pattern Recognition Workshops (CVPRW)},
  pages={217--226},
  year={2019},
  organization={IEEE},
  doi       = {10/ghm7jx}
}

@article{feng2025deep,
  title={Deep learning-based model for analyzing student engagement in activities},
  author={Feng, Feng},
  journal={Scientific Reports},
  year={2025},
  publisher={Nature Publishing Group UK London},
  doi       = {10/hbt7s3},
  volume    = {15},
  number    = {1}
}

@article{madsen2021synchronized,
  title={Synchronized eye movements predict test scores in online video education},
  author={Madsen, Jens and J{\'u}lio, Sara U and Gucik, Pawel J and Steinberg, Richard and Parra, Lucas C},
  journal={Proceedings of the National Academy of Sciences},
  volume={118},
  number={5},
  pages={e2016980118},
  year={2021},
  publisher={National Academy of Sciences},
  doi       = {10/gk2q5m}
}

@article{goldberg2021attentive,
  title={Attentive or not? Toward a machine learning approach to assessing students’ visible engagement in classroom instruction},
  author={Goldberg, Patricia and S{\"u}mer, {\"O}mer and St{\"u}rmer, Kathleen and Wagner, Wolfgang and G{\"o}llner, Richard and Gerjets, Peter and Kasneci, Enkelejda and Trautwein, Ulrich},
  journal={Educational Psychology Review},
  volume={33},
  number={1},
  pages={27--49},
  year={2021},
  publisher={Springer},
  doi       = {10/gghvzg}
}

@inproceedings{zhao2025research,
  title={Research on Classroom Behavior Analysis and Quantitative Evaluation System of Student Attention Based on Computer Vision},
  author={Zhao, Li and Sheng, Xinyu},
  booktitle={Proceedings of the 2025 6th International Conference on Computer Information and Big Data Applications},
  pages={1003--1008},
  year={2025},
  doi       = {10/hbt7s4}
}

@inproceedings{asish2023internal,
  title={Internal distraction detection utilizing eeg data in an educational vr environment},
  author={Asish, Sarker Monojit and Kulshreshth, Arun K and Borst, Christoph},
  booktitle={ACM Symposium on Applied Perception 2023},
  pages={1--10},
  year={2023},
  doi       = {10/hbt7s5}
}

@article{nair2024human,
  title={Human attention detection system using deep learning and brain--computer interface},
  author={Nair, S Anju Latha and Megalingam, Rajesh Kannan},
  journal={Neural Computing and Applications},
  volume={36},
  number={18},
  pages={10927--10940},
  year={2024},
  publisher={Springer},
  doi       = {10/hbt7s6}
}

@article{devi2024ga,
  title={GA-CNN: Analyzing student’s cognitive skills with EEG data using a hybrid deep learning approach},
  author={Devi, D and Sophia, S},
  journal={Biomedical Signal Processing and Control},
  volume={90},
  pages={105888},
  year={2024},
  publisher={Elsevier},
  doi       = {10/hbt7s7}
}

@article{rehman2025measuring,
  title={Measuring student attention based on EEG brain signals using deep reinforcement learning},
  author={Rehman, Asad Ur and Shi, Xiaochuan and Ullah, Farhan and Wang, Zepeng and Ma, Chao},
  journal={Expert Systems with Applications},
  volume={269},
  pages={126426},
  year={2025},
  publisher={Elsevier},
  doi       = {10/hbt7vj}
}

@article{yu2021reinforcement,
  title={Reinforcement learning in healthcare: A survey},
  author={Yu, Chao and Liu, Jiming and Nemati, Shamim and Yin, Guosheng},
  journal={ACM Computing Surveys (CSUR)},
  volume={55},
  number={1},
  pages={1--36},
  year={2021},
  publisher={ACM New York, NY},
  doi       = {10/grxn7q}
}

@inproceedings{blascheck2014state,
  title={State-of-the-art of visualization for eye tracking data.},
  author={Blascheck, Tanja and Kurzhals, Kuno and Raschke, Michael and Burch, Michael and Weiskopf, Daniel and Ertl, Thomas},
  booktitle={EuroVis - STARs},
  editor={R. Borgo and R. Maciejewski and I. Viola},
  pages={29},
  year={2014},
  publisher={The Eurographics Association},
  doi       = {10/gg689g}
}

@article{srinivasan2024attention,
  title={Attention-aware visualization: Tracking and responding to user perception over time},
  author={Srinivasan, Arvind and Ellemose, Johannes and Butcher, Peter WS and Ritsos, Panagiotis D and Elmqvist, Niklas},
  journal={IEEE Transactions on visualization and computer graphics},
  volume={31},
  number={1},
  pages={1017--1027},
  year={2024},
  publisher={IEEE},
  doi       = {10/hbqkwz}
}

@inproceedings{nguyen2015interactive,
  title={Interactive visualization for understanding of attention patterns},
  author={Nguyen, Truong-Huy D and El-Nasr, Magy Seif and Isaacowitz, Derek M},
  booktitle={Workshop on Eye Tracking and Visualization},
  pages={23--39},
  year={2015},
  organization={Springer},
  doi       = {10/hbt7tc}
}

@inproceedings{duchowski2015visualizing,
  title={Visualizing dynamic ambient/focal attention with coefficient},
  author={Duchowski, Andrew T and Krejtz, Krzysztof},
  booktitle={Workshop on Eye Tracking and Visualization},
  pages={217--233},
  year={2015},
  organization={Springer},
  doi       = {10/hbt7td}
}

@article{chang2025tell,
  title={Tell Me Without Telling Me: Two-Way Prediction of Visualization Literacy and Visual Attention},
  author={Chang, Minsuk and Wang, Yao and Wang, Huichen Will and Zhou, Yuanhong and Bulling, Andreas and Bearfield, Cindy Xiong},
  journal={IEEE Transactions on Visualization and Computer Graphics},
  year={2025},
  publisher={IEEE},
  doi       = {10/hbt7tf}
}

@inproceedings{jianu2025gaze,
  title={Gaze-Aware Visualisation: Design Considerations and Research Agenda},
  author={Jianu, Radu and Silva, Nelson and Rodrigues, Nils and Blascheck, Tanja and Schreck, Tobias and Weiskopf, Daniel},
  booktitle={Computer Graphics Forum},
  volume={44},
  number={6},
  pages={e70097},
  year={2025},
  organization={Wiley Online Library},
  doi       = {10/hbt7tg}
}

@article{conley2020examining,
  title={Examining course layouts in blackboard: Using eye-tracking to evaluate usability in a learning management system},
  author={Conley, Quincy and Earnshaw, Yvonne and McWatters, Grayley},
  journal={International Journal of Human--Computer Interaction},
  volume={36},
  number={4},
  pages={373--385},
  year={2020},
  publisher={Taylor \& Francis},
  doi       = {10/gh24hc}
}

@inproceedings{rahman2020exploring,
  title={Exploring eye gaze visualization techniques for identifying distracted students in educational VR},
  author={Rahman, Yitoshee and Asish, Sarker M and Fisher, Nicholas P and Bruce, Ethan C and Kulshreshth, Arun K and Borst, Christoph W},
  booktitle={2020 IEEE Conference on Virtual Reality and 3D User Interfaces (VR)},
  pages={868--877},
  year={2020},
  organization={IEEE},
  doi       = {10/ghvpgx}
}

@incollection{sauter2023behind,
  title={Behind the screens: Exploring eye movement visualization to optimize online teaching and learning},
  author={Sauter, Marian and Wagner, Tobias and Hirzle, Teresa and Lin, Bao Xin and Rukzio, Enrico and Huckauf, Anke},
  booktitle={Proceedings of Mensch und Computer 2023},
  pages={67--80},
  year={2023},
  doi       = {10/hbt7tk},
  publisher = {ACM}
}

@inproceedings{hirzle2022attention,
  title={Attention of many observers visualized by eye movements},
  author={Hirzle, Teresa and Sauter, Marian and Wagner, Tobias and Hummel, Susanne and Rukzio, Enrico and Huckauf, Anke},
  booktitle={2022 Symposium on eye tracking research and applications},
  pages={1--7},
  year={2022},
  doi       = {10/grp6kc}
}

@inproceedings{davalos2024gazeviz,
  title={GazeViz: A web-based approach for visualizing learner gaze patterns in online educational environment},
  author={Davalos, Eduardo and Srivastava, Namrata and Zhang, Yike and Goodwin, Amanda and Biswas, Gautam},
  booktitle={International Conference on Computers in Education},
  year={2024},
  doi       = {10/hbt7tm}
}

@inproceedings{thanyadit2023tutor,
  title={Tutor in-sight: Guiding and visualizing students’ attention with mixed reality avatar presentation tools},
  author={Thanyadit, Santawat and Heintz, Matthias and Law, Effie LC},
  booktitle={Proceedings of the 2023 CHI Conference on Human Factors in Computing Systems},
  pages={1--20},
  year={2023},
  doi       = {10/gr7h28}
}

@article{liu2024eeg,
  title={An EEG motor imagery dataset for brain computer interface in acute stroke patients},
  author={Liu, Haijie and Wei, Penghu and Wang, Haochong and Lv, Xiaodong and Duan, Wei and Li, Meijie and Zhao, Yan and Wang, Qingmei and Chen, Xinyuan and Shi, Gaige and others},
  journal={Scientific Data},
  volume={11},
  number={1},
  pages={131},
  year={2024},
  publisher={Nature Publishing Group UK London},
  doi       = {10/g88cwc}
}

@article{fazli2012enhanced,
  title={Enhanced performance by a hybrid NIRS--EEG brain computer interface},
  author={Fazli, Siamac and Mehnert, Jan and Steinbrink, Jens and Curio, Gabriel and Villringer, Arno and M{\"u}ller, Klaus-Robert and Blankertz, Benjamin},
  journal={Neuroimage},
  volume={59},
  number={1},
  pages={519--529},
  year={2012},
  publisher={Elsevier},
  doi       = {10/dzg3t3}
}

@article{mnih2015human,
  title={Human-level control through deep reinforcement learning},
  author={Mnih, Volodymyr and Kavukcuoglu, Koray and Silver, David and Rusu, Andrei A and Veness, Joel and Bellemare, Marc G and Graves, Alex and Riedmiller, Martin and Fidjeland, Andreas K and Ostrovski, Georg and others},
  journal={nature},
  volume={518},
  number={7540},
  pages={529--533},
  year={2015},
  publisher={Nature Publishing Group},
  doi       = {10/gc3h75}
}

@inproceedings{zhao2025wearable,
 title={Wearable Music2Emotion: Assessing Emotions Induced by AI-Generated Music through Portable EEG-fNIRS Fusion},
 author={Zhao, Sha and Yi, Song and Zhou, Yangxuan and Pan, Jiadong and Wang, Jiquan and Xia, Jie and Li, Shijian and Dong, Shurong and Pan, Gang},
 booktitle={Proceedings of the 33rd ACM International Conference on Multimedia},
 pages={5627--5636},
 year={2025},
  doi       = {10/hbt7tq}
}

@article{corbetta2002control,
  title={Control of goal-directed and stimulus-driven attention in the brain},
  author={Corbetta, Maurizio and Shulman, Gordon L},
  journal={Nature reviews neuroscience},
  volume={3},
  number={3},
  pages={201--215},
  year={2002},
  publisher={Nature Publishing Group UK London},
  doi       = {10/brm459}
}

@article{gregoriou2009high,
  title={High-frequency, long-range coupling between prefrontal and visual cortex during attention},
  author={Gregoriou, Georgia G and Gotts, Stephen J and Zhou, Huihui and Desimone, Robert},
  journal={science},
  volume={324},
  number={5931},
  pages={1207--1210},
  year={2009},
  publisher={American Association for the Advancement of Science},
  doi       = {10/ckh48g}
}

@article{shatil2014novel,
  title={Novel television-based cognitive training improves working memory and executive function},
  author={Shatil, Evelyn and Mikulecka, Jaroslava and Bellotti, Francesco and Bure{\v{s}}, Vladim{\'\i}r},
  journal={PloS one},
  volume={9},
  number={7},
  pages={e101472},
  year={2014},
  publisher={Public Library of Science San Francisco, USA},
  doi       = {10/gf9jff}
}

@article{aci2019distinguishing,
  title={Distinguishing mental attention states of humans via an EEG-based passive BCI using machine learning methods},
  author={Ac{\i}, {\c{C}}i{\u{g}}dem {\.I}nan and Kaya, Murat and Mishchenko, Yuriy},
  journal={Expert Systems with Applications},
  volume={134},
  pages={153--166},
  year={2019},
  publisher={Elsevier},
  doi       = {10/ggzbr2}
}

@article{gao2023hybrid,
  title={Hybrid EEG-fNIRS brain computer interface based on common spatial pattern by using EEG-informed general linear model},
  author={Gao, Yunyuan and Jia, Biao and Houston, Michael and Zhang, Yingchun},
  journal={IEEE Transactions on Instrumentation and Measurement},
  volume={72},
  pages={1--10},
  year={2023},
  publisher={IEEE},
  doi       = {10/hbt7ts}
}

@article{desai2021generalizable,
  title={Generalizable EEG encoding models with naturalistic audiovisual stimuli},
  author={Desai, Maansi and Holder, Jade and Villarreal, Cassandra and Clark, Nat and Hoang, Brittany and Hamilton, Liberty S},
  journal={Journal of Neuroscience},
  volume={41},
  number={43},
  pages={8946--8962},
  year={2021},
  publisher={Society for Neuroscience},
  doi       = {10/gjp3f3}
}

@article{fisher2018limited,
  title={The limited capacity model of motivated mediated message processing: Taking stock of the past},
  author={Fisher, Jacob T and Keene, Justin Robert and Huskey, Richard and Weber, Ren{\'e}},
  journal={Annals of the International Communication Association},
  volume={42},
  number={4},
  pages={270--290},
  year={2018},
  publisher={Taylor \& Francis},
  doi       = {10/gmgqx8}
}

@article{rosenholtz2007measuring,
  title={Measuring visual clutter},
  author={Rosenholtz, Ruth and Li, Yuanzhen and Nakano, Lisa},
  journal={Journal of vision},
  volume={7},
  number={2},
  pages={17--17},
  year={2007},
  publisher={The Association for Research in Vision and Ophthalmology},
  doi       = {10/bqtpr4}
}

@article{shannon1948mathematical,
  title={A mathematical theory of communication},
  author={Shannon, Claude Elwood},
  journal={The Bell system technical journal},
  volume={27},
  number={3},
  pages={379--423},
  year={1948},
  publisher={Nokia Bell Labs},
  doi       = {10/b39t}
}

@article{apicella2022eeg,
  title={EEG-based measurement system for monitoring student engagement in learning 4.0},
  author={Apicella, Andrea and Arpaia, Pasquale and Frosolone, Mirco and Improta, Giovanni and Moccaldi, Nicola and Pollastro, Andrea},
  journal={Scientific Reports},
  volume={12},
  number={1},
  pages={5857},
  year={2022},
  publisher={Nature Publishing Group UK London},
  doi       = {10/hbc6nw}
}

@article{zhang2023unsupervised,
  title={Unsupervised time-aware sampling network with deep reinforcement learning for EEG-based emotion recognition},
  author={Zhang, Yongtao and Pan, Yue and Zhang, Yulin and Zhang, Min and Li, Linling and Zhang, Li and Huang, Gan and Su, Lei and Liu, Honghai and Liang, Zhen and others},
  journal={IEEE Transactions on Affective Computing},
  volume={15},
  number={3},
  pages={1090--1103},
  year={2023},
  publisher={IEEE},
  doi       = {10/g7zrw2}
}

@incollection{shneiderman2003eyes,
  title={The eyes have it: A task by data type taxonomy for information visualizations},
  author={Shneiderman, Ben},
  booktitle={The craft of information visualization},
  pages={364--371},
  year={2003},
  publisher={Elsevier},
  doi       = {10/fr3bv2}
}

@article{hager2021biomarker,
  title={Biomarker support for ADHD diagnosis based on event related potentials and scores from an attention test},
  author={H{\"a}ger, LA and Johnels, J {\AA}sberg and Kropotov, JD and Weidle, B and Hollup, S and Zehentbauer, PG and Gillberg, C and Billstedt, E and Ogrim, G},
  journal={Psychiatry research},
  volume={300},
  pages={113879},
  year={2021},
  publisher={Elsevier},
  doi       = {10/gmkj5c}
}

@article{pellinen2024improving,
  title={Improving epilepsy diagnosis across the lifespan: approaches and innovations},
  author={Pellinen, Jacob and Foster, Emma C and Wilmshurst, Jo M and Zuberi, Sameer M and French, Jacqueline},
  journal={The Lancet Neurology},
  volume={23},
  number={5},
  pages={511--521},
  year={2024},
  publisher={Elsevier},
  doi       = {10/hbt7t2}
}

@article{chen2017assessing,
  title={Assessing the attention levels of students by using a novel attention aware system based on brainwave signals},
  author={Chen, Chih-Ming and Wang, Jung-Ying and Yu, Chih-Ming},
  journal={British Journal of Educational Technology},
  volume={48},
  number={2},
  pages={348--369},
  year={2017},
  publisher={Wiley Online Library},
  doi       = {10/f9r3h7}
}

@inproceedings{wen2023nftdisk,
author = {Wen, Xiaolin and Wang, Yong and Yue, Xuanwu and Zhu, Feida and Zhu, Min},
title = {NFTDisk: Visual Detection of Wash Trading in NFT Markets},
year = {2023},
isbn = {9781450394215},
publisher = {Association for Computing Machinery},
address = {New York, NY, USA},
url = {https://doi.org/10.1145/3544548.3581466},
doi = {10.1145/3544548.3581466},
booktitle = {Proceedings of the 2023 CHI Conference on Human Factors in Computing Systems},
articleno = {215},
numpages = {15},
location = {Hamburg, Germany},
series = {CHI '23}
}

@inproceedings{he2017mask,
  title={Mask r-cnn},
  author={He, Kaiming and Gkioxari, Georgia and Doll{\'a}r, Piotr and Girshick, Ross},
  booktitle={Proceedings of the IEEE international conference on computer vision},
  pages={2961--2969},
  year={2017}
}

@inproceedings{liao2020real,
  title={Real-time scene text detection with differentiable binarization},
  author={Liao, Minghui and Wan, Zhaoyi and Yao, Cong and Chen, Kai and Bai, Xiang},
  booktitle={Proceedings of the AAAI conference on artificial intelligence},
  volume={34},
  number={07},
  pages={11474--11481},
  year={2020}
}

@article{attneave1954some,
  title={Some informational aspects of visual perception.},
  author={Attneave, Fred},
  journal={Psychological review},
  volume={61},
  number={3},
  pages={183},
  year={1954},
  publisher={American Psychological Association}
}

@article{donderi2006visual,
  title={Visual complexity: a review.},
  author={Donderi, Don C},
  journal={Psychological bulletin},
  volume={132},
  number={1},
  pages={73},
  year={2006},
  publisher={American Psychological Association}
}

@inproceedings{miniukovich2014quantification,
  title={Quantification of interface visual complexity},
  author={Miniukovich, Aliaksei and De Angeli, Antonella},
  booktitle={Proceedings of the 2014 international working conference on advanced visual interfaces},
  pages={153--160},
  year={2014}
}

@inproceedings{reinecke2013,
  title={Predicting users' first impressions of website aesthetics with a quantification of perceived visual complexity and colorfulness},
  author={Reinecke, Katharina and Yeh, Tom and Miratrix, Luke and Mardiko, Rahmatri and Zhao, Yuechen and Liu, Jenny and Gajos, Krzysztof Z},
  booktitle={Proceedings of the SIGCHI conference on human factors in computing systems},
  pages={2049--2058},
  year={2013}
}

@inproceedings{shi2019,
  title={Investigating correlations of automatically extracted multimodal features and lecture video quality},
  author={Shi, Jianwei and Otto, Christian and Hoppe, Anett and Holtz, Peter and Ewerth, Ralph},
  booktitle={Proceedings of the 1st international workshop on search as learning with multimedia information},
  pages={11--19},
  year={2019}
}

@article{ferguson2017,
  title={Continuing medical education speakers with high evaluation scores use more image-based slides},
  author={Ferguson, Ian and Phillips, Andrew W and Lin, Michelle},
  journal={Western Journal of Emergency Medicine},
  volume={18},
  number={1},
  pages={152},
  year={2016}
}

@article{shannon1951prediction,
  title={Prediction and entropy of printed English},
  author={Shannon, Claude E},
  journal={Bell system technical journal},
  volume={30},
  number={1},
  pages={50--64},
  year={1951},
  publisher={Wiley Online Library}
}

@article{bentz2017entropy,
  title={The entropy of words—Learnability and expressivity across more than 1000 languages},
  author={Bentz, Christian and Alikaniotis, Dimitrios and Cysouw, Michael and Ferrer-i-Cancho, Ramon},
  journal={Entropy},
  volume={19},
  number={6},
  pages={275},
  year={2017},
  publisher={MDPI}
}

@article{geirnaert2021,
  title={Electroencephalography-based auditory attention decoding: Toward neurosteered hearing devices},
  author={Geirnaert, Simon and Vandecappelle, Servaas and Alickovic, Emina and De Cheveigne, Alain and Lalor, Edmund and Meyer, Bernd T and Miran, Sina and Francart, Tom and Bertrand, Alexander},
  journal={IEEE Signal Processing Magazine},
  volume={38},
  number={4},
  pages={89--102},
  year={2021},
  publisher={IEEE}
}

@article{ahn2025drives,
  title={What Drives Student Engagement and Learning in Video Lectures? An Investigation of Instructor Visibility, Playback Speed, and Student Preferences},
  author={Ahn, Dahwi and Chan, Jason CK},
  journal={Applied Cognitive Psychology},
  volume={39},
  number={2},
  pages={e70026},
  year={2025},
  publisher={Wiley Online Library}
}

@article{beauchamp2004integration,
  title={Integration of auditory and visual information about objects in superior temporal sulcus},
  author={Beauchamp, Michael S and Lee, Kathryn E and Argall, Brenna D and Martin, Alex},
  journal={Neuron},
  volume={41},
  number={5},
  pages={809--823},
  year={2004},
  publisher={Elsevier}
}

@article{noppeney2021,
  title={Perceptual inference, learning, and attention in a multisensory world},
  author={Noppeney, Uta},
  journal={Annual review of neuroscience},
  volume={44},
  number={1},
  pages={449--473},
  year={2021},
  publisher={Annual Reviews}
}

@article{grill2004human,
  title={The human visual cortex},
  author={Grill-Spector, Kalanit and Malach, Rafael},
  journal={Annu. Rev. Neurosci.},
  volume={27},
  number={1},
  pages={649--677},
  year={2004},
  publisher={Annual Reviews}
}

@article{esterman2013zone,
  title={In the zone or zoning out? Tracking behavioral and neural fluctuations during sustained attention},
  author={Esterman, Michael and Noonan, Sarah K and Rosenberg, Monica and DeGutis, Joseph},
  journal={Cerebral cortex},
  volume={23},
  number={11},
  pages={2712--2723},
  year={2013},
  publisher={Oxford University Press}
}

@article{berka2007eeg,
  title={EEG correlates of task engagement and mental workload in vigilance, learning, and memory tasks},
  author={Berka, Chris and Levendowski, Daniel J and Lumicao, Michelle N and Yau, Alan and Davis, Gene and Zivkovic, Vladimir T and Olmstead, Richard E and Tremoulet, Patrice D and Craven, Patrick L},
  journal={Aviation, space, and environmental medicine},
  volume={78},
  number={5},
  pages={B231--B244},
  year={2007},
  publisher={Aerospace Medical Association}
}

@article{larson2017sample,
  title={Sample size calculations in human electrophysiology (EEG and ERP) studies: A systematic review and recommendations for increased rigor},
  author={Larson, Michael J and Carbine, Kaylie A},
  journal={International Journal of Psychophysiology},
  volume={111},
  pages={33--41},
  year={2017},
  publisher={Elsevier}
}

@article{mcvay2012drifting,
  title={Drifting from slow to “d'oh!”: Working memory capacity and mind wandering predict extreme reaction times and executive control errors.},
  author={McVay, Jennifer C and Kane, Michael J},
  journal={Journal of Experimental Psychology: Learning, Memory, and Cognition},
  volume={38},
  number={3},
  pages={525},
  year={2012},
  publisher={American Psychological Association}
}

@article{fishburn2014sensitivity,
  title={Sensitivity of fNIRS to cognitive state and load},
  author={Fishburn, Frank A and Norr, Megan E and Medvedev, Andrei V and Vaidya, Chandan J},
  journal={Frontiers in human neuroscience},
  volume={8},
  pages={76},
  year={2014},
  publisher={Frontiers Media SA}
}

@article{warrier2009relating,
  title={Relating structure to function: Heschl's gyrus and acoustic processing},
  author={Warrier, Catherine and Wong, Patrick and Penhune, Virginia and Zatorre, Robert and Parrish, Todd and Abrams, Daniel and Kraus, Nina},
  journal={Journal of Neuroscience},
  volume={29},
  number={1},
  pages={61--69},
  year={2009},
  publisher={Society for Neuroscience}
}

@article{michel2012towards,
  title={Towards the utilization of {EEG} as a brain imaging tool},
  author={Michel, Christoph M and Murray, Micah M},
  journal={NeuroImage},
  volume={61}, number={2}, pages={371--385}, year={2012}, publisher={Elsevier}
}

@article{srinivasan1998spatial,
  title={Spatial filtering and neocortical dynamics: estimates of {EEG} coherence},
  author={Srinivasan, Ramesh and Nunez, Paul L and Silberstein, Richard B},
  journal={IEEE Transactions on Biomedical Engineering},
  volume={45}, number={7}, pages={814--826}, year={1998}, publisher={IEEE}
}

@book{nunez2006electric,
  title={Electric Fields of the Brain: The Neurophysics of {EEG}},
  author={Nunez, Paul L and Srinivasan, Ramesh},
  edition={2nd}, year={2006}, publisher={Oxford University Press}
}

@article{grech2008review,
  title={Review on solving the inverse problem in {EEG} source analysis},
  author={Grech, Roberta and Cassar, Tracey and Muscat, Joseph and Camilleri, Kenneth P and Fabri, Simon G and Zervakis, Michalis and Xanthopoulos, Petros and Sakkalis, Vangelis and Vanrumste, Bart},
  journal={Journal of NeuroEngineering and Rehabilitation},
  volume={5}, number={1}, pages={25}, year={2008}, publisher={Springer}
}

@article{ferrari2012brief,
  title={A brief review on the history of human functional near-infrared spectroscopy ({fNIRS}) development and fields of application},
  author={Ferrari, Marco and Quaresima, Valentina},
  journal={NeuroImage},
  volume={63}, number={2}, pages={921--935}, year={2012}, publisher={Elsevier}
}

@article{mayer2003nine,
  title={Nine ways to reduce cognitive load in multimedia learning},
  author={Mayer, Richard E and Moreno, Roxana},
  journal={Educational Psychologist},
  volume={38}, number={1}, pages={43--52}, year={2003}, publisher={Taylor \& Francis}
}

@article{lim2010meta,
  title={A meta-analysis of the impact of short-term sleep deprivation on cognitive variables},
  author={Lim, Julian and Dinges, David F},
  journal={Psychological Bulletin},
  volume={136}, number={3}, pages={375}, year={2010}, publisher={American Psychological Association}
}

@article{kalyuga2007expertise,
  title={Expertise reversal effect and its implications for learner-tailored instruction},
  author={Kalyuga, Slava},
  journal={Educational Psychology Review},
  volume={19}, number={4}, pages={509--539}, year={2007}, publisher={Springer}
}

@article{spearman1961proof,
  title={The proof and measurement of association between two things.},
  author={Spearman, Charles},
  year={1961},
  publisher={Appleton-Century-Crofts}
}

@article{jaccard1912distribution,
  title={The distribution of the flora in the alpine zone},
  author={Jaccard, Paul},
  journal={New Phytologist}, volume={11}, number={2}, pages={37--50}, year={1912}, publisher={Wiley}
}

@article{kendall1938new,
  title={A new measure of rank correlation},
  author={Kendall, Maurice G},
  journal={Biometrika}, volume={30}, number={1/2}, pages={81--93}, year={1938}
}

@article{coupe2019different,
  title={Different languages, similar encoding efficiency: Comparable information rates across the human communicative niche},
  author={Coup{\'e}, Christophe and Oh, Yoon Mi and Dediu, Dan and Pellegrino, Fran{\c{c}}ois},
  journal={Science advances},
  volume={5},
  number={9},
  pages={eaaw2594},
  year={2019},
  publisher={American Association for the Advancement of Science}
}

@article{mccarthy2010mtld,
  title={MTLD, vocd-D, and HD-D: A validation study of sophisticated approaches to lexical diversity assessment},
  author={McCarthy, Philip M and Jarvis, Scott},
  journal={Behavior research methods},
  volume={42},
  number={2},
  pages={381--392},
  year={2010},
  publisher={Springer}
}

@article{nothelfer2017redundant,
  title={Redundant encoding strengthens segmentation and grouping in visual displays of data.},
  author={Nothelfer, Christine and Gleicher, Michael and Franconeri, Steven},
  journal={Journal of Experimental Psychology: Human Perception and Performance},
  volume={43},
  number={9},
  pages={1667},
  year={2017},
  publisher={American Psychological Association}
}

@article{chun2017redundant,
  title={Redundant encoding in data visualizations: Assessing perceptual accuracy and speed},
  author={Chun, Russell},
  journal={Visual Communication Quarterly},
  volume={24},
  number={3},
  pages={135--148},
  year={2017},
  publisher={Taylor \& Francis}
}

@article{nesbit2006learning,
  title={Learning with concept and knowledge maps: A meta-analysis},
  author={Nesbit, John C and Adesope, Olusola O},
  journal={Review of educational research},
  volume={76},
  number={3},
  pages={413--448},
  year={2006},
  publisher={Sage Publications Sage CA: Thousand Oaks, CA}
}

@article{brehmer2016timelines,
  title={Timelines revisited: A design space and considerations for expressive storytelling},
  author={Brehmer, Matthew and Lee, Bongshin and Bach, Benjamin and Riche, Nathalie Henry and Munzner, Tamara},
  journal={IEEE transactions on visualization and computer graphics},
  volume={23},
  number={9},
  pages={2151--2164},
  year={2016},
  publisher={IEEE}
}

@article{wan2021frontal,
  title={Frontal EEG-based multi-level attention states recognition using dynamical complexity and extreme gradient boosting},
  author={Wan, Wang and Cui, Xingran and Gao, Zhilin and Gu, Zhongze},
  journal={Frontiers in human neuroscience},
  volume={15},
  pages={673955},
  year={2021},
  publisher={Frontiers Media SA}
}
 
\begin{IEEEbiography}[{\includegraphics[width=1in,height=1.25in,clip,keepaspectratio]{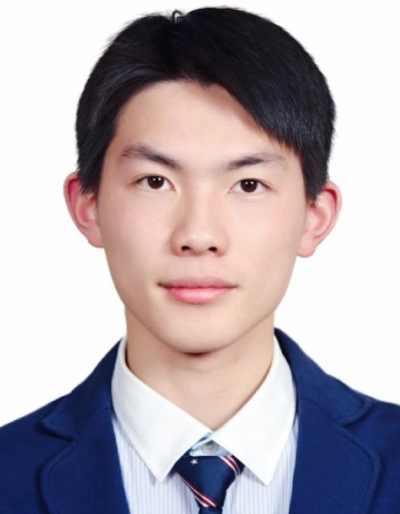}}]{Shixian Zhou}
  is currently working toward the MPhil degree with the School of Media and Design, Hangzhou Dianzi University, Hangzhou, China. He also studied as a visiting student at Nanyang Technological University, Singapore. His research interests include visual analytics, brain-computer interfaces, and online learning data analysis.
\end{IEEEbiography}

\begin{IEEEbiography}[{\includegraphics[width=1in,height=1.25in,clip,keepaspectratio]{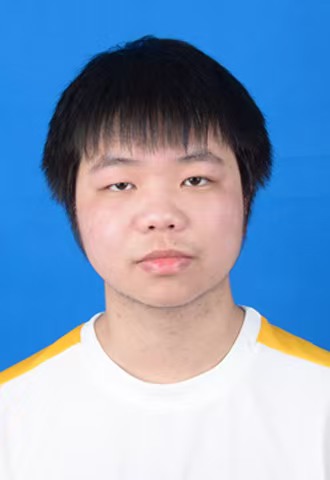}}]{Minghuan Shen}
  is currently working toward the postgraduate degree at Hangzhou Dianzi University. His research interests include data visualization and explainable machine learning.
\end{IEEEbiography}

\begin{IEEEbiography}[{\includegraphics[width=1in,height=1.25in,clip,keepaspectratio]{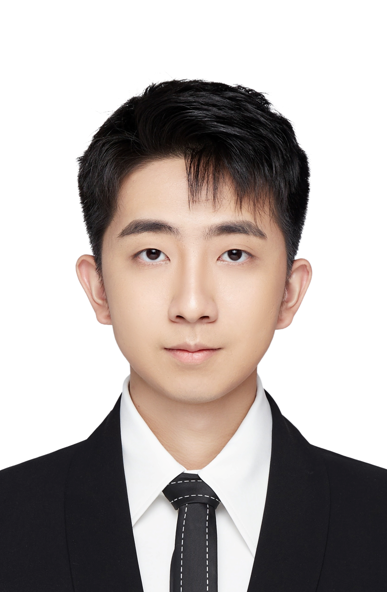}}]{Xiaolin Wen}
  is currently a Ph.D student in the College of Computing and Data Science, Nanyang Technological University (NTU).
  His research interests mainly focus on visualization for FinTech and LLM-assisted design study.
  He received his master's degree in Computer Science and Technology from Sichuan University in 2023 and his dual bachelor's degree in Computer Science and Financial Engineering from Sichuan University in 2016.
  For more information, kindly visit \url{https://wenxiaolin.com/}.
\end{IEEEbiography}

\begin{IEEEbiography}[{\includegraphics[width=1in,height=1.25in,clip,keepaspectratio]{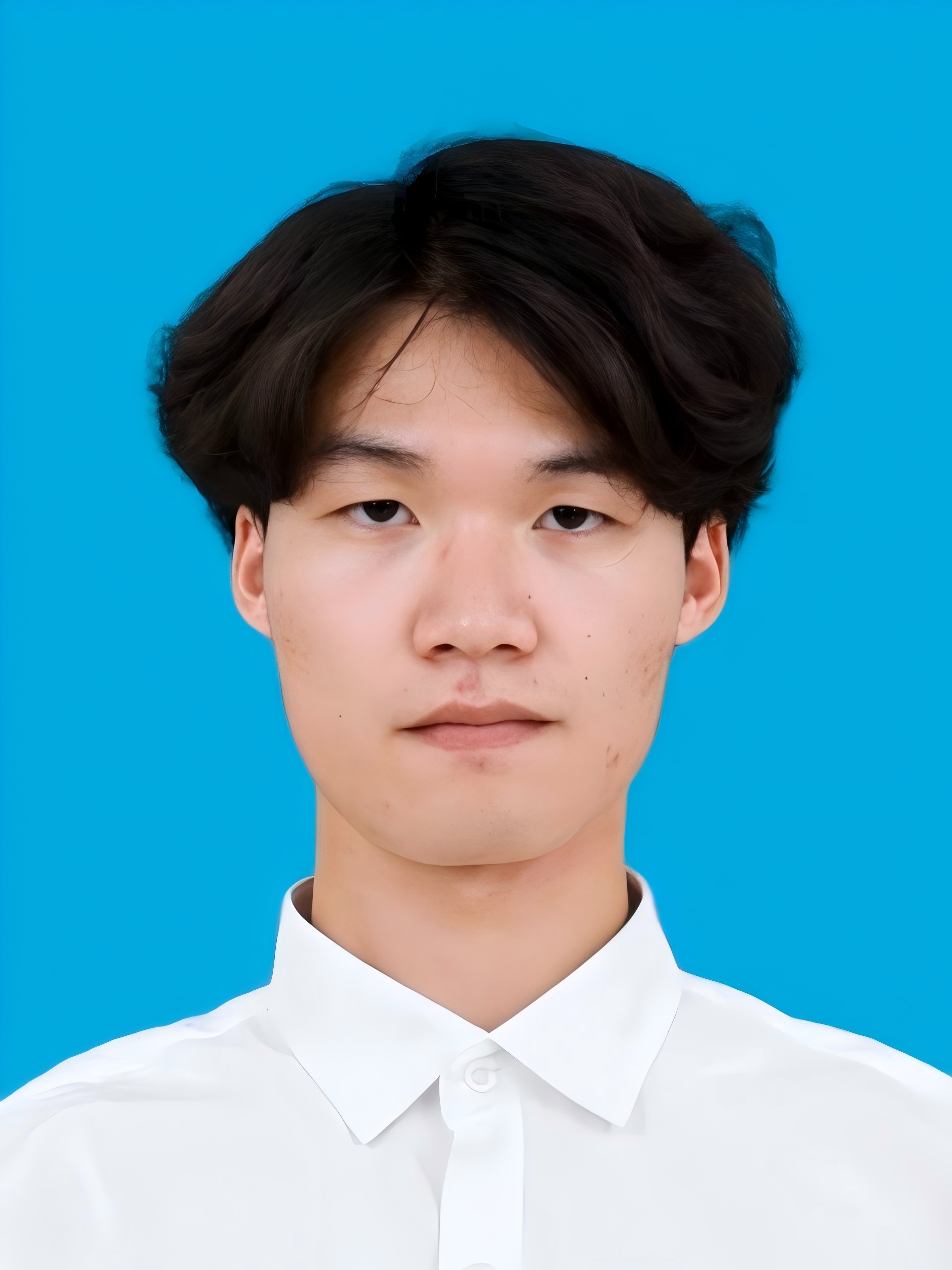}}]{Zijun Qiu}
  is currently pursuing the M.S. degree in Digital Media Technology at Hangzhou Dianzi University, Hangzhou, China. His research interests include data visualization and visual analytics.
\end{IEEEbiography}

\begin{IEEEbiography}[{\includegraphics[width=1in,height=1.25in,clip,keepaspectratio]{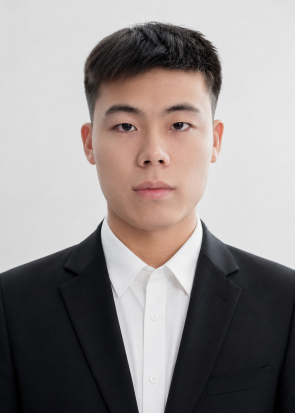}}]{Yongliang Jiang}
  is currently pursuing a postgraduate degree at Hangzhou Dianzi University. His research interests include visual analytics and human-computer interaction.
\end{IEEEbiography}

\begin{IEEEbiography}[{\includegraphics[width=1in,height=1.25in,clip,keepaspectratio]{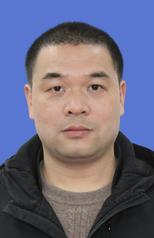}}]{Xiangyang Wu}
received the PhD degree in math from
Zhejiang University, China, in 2006. He is currently
a professor with the School of Computer Science,
Hangzhou Dianzi University. His current research
interests include large-scale data analytics, data visualization and visual analytics, transportation network
modeling, etc.
\end{IEEEbiography}

\begin{IEEEbiography}[{\includegraphics[width=1in,height=1.25in,clip,keepaspectratio]{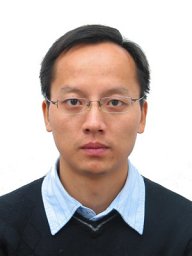}}]{Fei Wu}
 (Senior Member, IEEE) received the Ph.D. degree from Zhejiang University, Hangzhou, China, in 2002.,He was a Visiting Scholar with Prof. B. Yu’s Group, University of California at Berkeley, Berkeley, CA, USA, from 2009 to 2010. He is currently a Full Professor with the College of Computer Science and Technology, Zhejiang University. His current research interests include multimedia retrieval, sparse representation, and machine learning.
\end{IEEEbiography}

\begin{IEEEbiography}[{\includegraphics[width=1.0in,height=1.25in,clip,keepaspectratio]{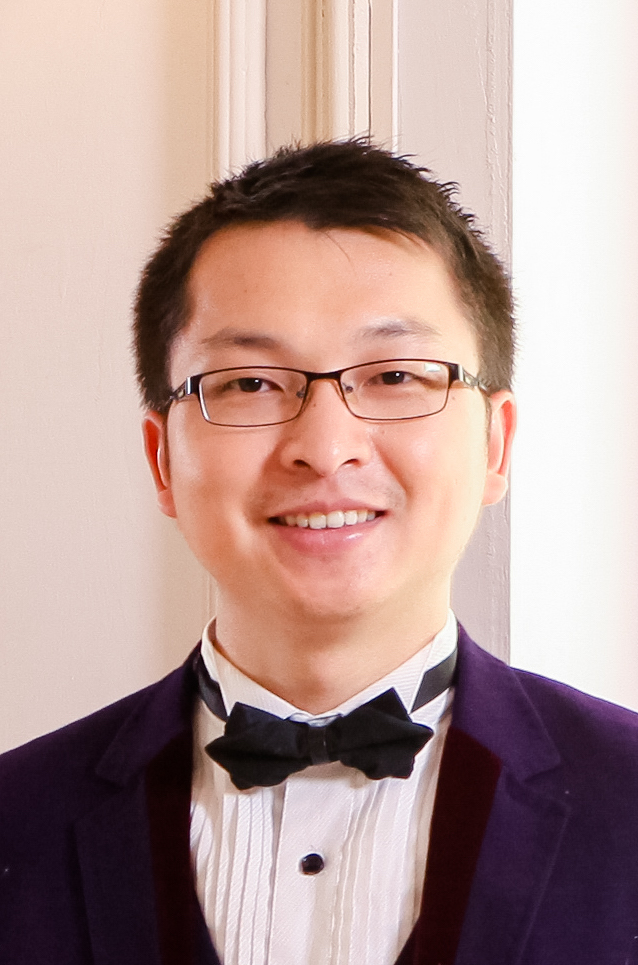}}]{Yong Wang}
is currently an assistant professor in the College of Computing and Data Science, Nanyang Technological University. Before that, he worked as an assistant professor at Singapore Management University from 2020 to 2024. His research interests include information visualization, visual analytics and human-AI collaboration, with an emphasis on their application to FinTech, quantum computing and online learning. He obtained his Ph.D. in Computer Science from Hong Kong University of Science and Technology. He received his B.E. and M.E. from Harbin Institute of Technology and Huazhong University of Science and Technology, respectively. For more details, please refer to \url{http://yong-wang.org}.
\end{IEEEbiography}

\begin{IEEEbiography}[{\includegraphics[width=1in,height=1.25in,clip,keepaspectratio]{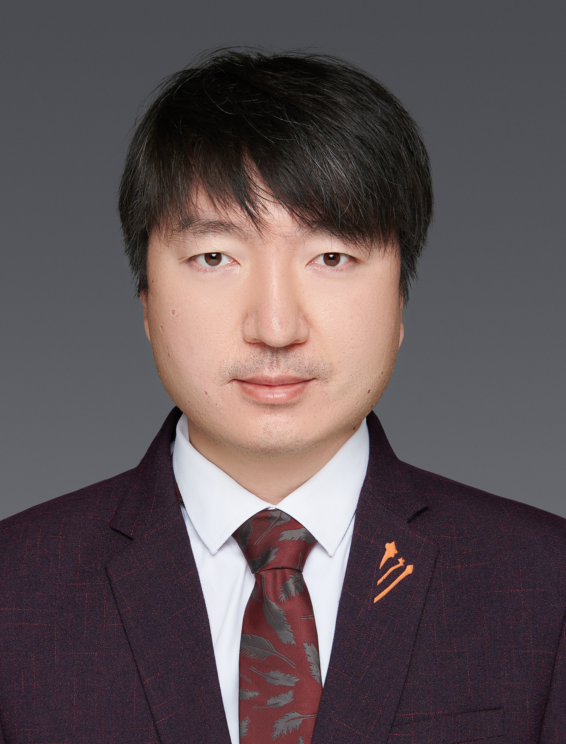}}]{Zhiguang Zhou}
is currently a professor in School of Media and Design and serves as the dean of Digital Media Technology Research Institute at Hangzhou Dianzi University. His research interests include data visualization, visual analytics and knowledge graph mining. He received his Ph.D. in Computer Science from the state key Laboratory of CAD\&CG in Zhejiang University. 
\end{IEEEbiography}

\vfill

\end{document}